\documentclass[aps,prd,longbibliography,preprintnumbers,nofootinbib,
superscriptaddress,amsmath,floatfix]{revtex4}

\usepackage{amssymb}  
\usepackage{graphicx,subfigure} 
\usepackage{exscale}
\usepackage{textcomp}
\usepackage{enumerate}
\usepackage [english]{babel}
\usepackage[utf8]{inputenc}
\usepackage [autostyle, english = american]{csquotes}
\usepackage{slashed}
\usepackage{hyperref}

\usepackage{color}

\usepackage[normalem]{ulem}  

\newcommand{\non}{\nonumber\\}

\newcommand{\be}{\begin{equation}}
\newcommand{\ee}{\end{equation}}
\newcommand{\bea}{\begin{eqnarray}}
\newcommand{\eea}{\end{eqnarray}}
\newcommand{\ba}[1]{\begin{array}{#1}}
\newcommand{\ea}{\end{array}}

\newcommand{\Tr}{{\rm Tr}}

\definecolor{darkgreen}{rgb}{0.0,0.5,0}

\begin{document}

\title{Strange  quark matter from a baryonic approach}

\author{Eduardo S.\ Fraga}
\email{fraga@if.ufrj.br}
\affiliation{Instituto de F\'{\i}sica, Universidade Federal do Rio de Janeiro, 
Caixa Postal 68528, 21941-972, Rio de Janeiro, RJ, Brazil}

\author{Rodrigo da Mata}
\email{rsilva@pos.if.ufrj.br}
\affiliation{Instituto de F\'{\i}sica, Universidade Federal do Rio de Janeiro, 
Caixa Postal 68528, 21941-972, Rio de Janeiro, RJ, Brazil}

\author{Savvas Pitsinigkos}
\email{S.Pitsinigkos@soton.ac.uk}
\affiliation{Mathematical Sciences and STAG Research Centre, University of Southampton, Southampton SO17 1BJ, United Kingdom}

\author{Andreas Schmitt}
\email{a.schmitt@soton.ac.uk}
\affiliation{Mathematical Sciences and STAG Research Centre, University of Southampton, Southampton SO17 1BJ, United Kingdom}

\date{27 September 2022}

\begin{abstract} 
We construct a model for dense matter based on low-density nuclear matter properties that exhibits a chiral phase transition and that includes strangeness through hyperonic degrees of freedom. Empirical constraints from nuclear matter alone allow for various scenarios,  from a strong first-order chiral transition at relatively low densities through a weaker transition at higher densities, even up to a smooth crossover not far beyond the edge of the allowed range. The model parameters can be chosen such that at asymptotically large densities the chirally restored phase contains strangeness and the speed of sound  approaches the conformal limit, resulting in a high-density phase that resembles deconfined quark matter. Additionally, if the model is required to reproduce sufficiently massive compact stars, the allowed parameter range is significantly narrowed down, resulting for instance in a very narrow range for the poorly known slope parameter of the symmetry energy, $L\simeq (88-92)\, {\rm MeV}$. We also find that for the allowed parameter range strangeness does not appear in the form of hyperons in the chirally broken phase and the chiral transition is of first order. Due to its unified approach and relative simplicity -- here we restrict ourselves to zero temperature and the mean-field approximation -- the model can be used in the future to study dense matter under compact  star conditions in the vicinity of the chiral phase transition, for instance to compute the surface tension or to investigate spatially inhomogeneous phases. 
\end{abstract}

\maketitle


\section{Introduction}

Hadronic matter is expected to undergo a transition to 
the quark-gluon plasma at sufficiently large temperatures or baryon densities. The nature of this transition and its 
location in the phase diagram is poorly known, except for very small baryon chemical potentials, where a smooth crossover is predicted from lattice calculations \cite{Aoki:2006br}. In this paper, we are interested in the transition -- here: the chiral phase transition -- at zero temperature and thus large, but not asymptotically large,  baryon chemical potentials, a region not accessible from first principles with currently available methods. Besides its significance for the QCD phase diagram, this transition is highly relevant for the properties of compact stars, whose interiors may host deconfined quark matter \cite{Annala:2019puf}.

We do not intend to describe this transition from first principles, we shall rather employ a relatively simple phenomenological model. Even giving up the rigor of the underlying fundamental theory, it is a challenge to account for both quark matter and hadronic matter within a single approach. 
Using a single model is beneficial if one is interested for example in the critical chemical potential at which the chiral phase transition occurs. While this transition point is a prediction of a unified approach, it is essentially a model parameter if two separate descriptions of hadronic and quark matter are glued together. Another advantage is that properties in the vicinity of the phase transition can be calculated more reliably, most notably the surface tension (if the transition is of first order), for which the full potential, connecting both local minima, needs to be known\footnote{Even though there is a well-known systematic procedure to build an effective potential from the matching of pressures obtained from different models, it requires extra information that is usually not available, which results in more free parameters and uncertainties \cite{PhysRevA.8.3230,PhysRevA.22.2189,Csernai:1992tj}. In particular, one needs information on the barrier, which is directly related to the surface tension.}. 
The reason for that lies in the fact that, given the effective potential, one can resort to the full power of semiclassical methods, perturbing the system around classical solutions instead of trivial vacua (see e.g.\ Refs. \cite{Coleman:1977py,Callan:1977pt}). Such classical solutions probe the entire structure of the potential, including the different phases it allows.  

The majority of studies of hybrid stars -- compact stars with a quark matter core and a nuclear mantle -- employ two separate descriptions for the two phases, see for instance Refs.\ \cite{Alford:2013aca,Christian:2018jyd,Irving:2019yae,Pereira:2020jgv,Blaschke:2020qqj,Jokela:2020piw,Ferreira:2020kvu,Ferreira:2021osk,Lopes:2021jpm}. A few unified approaches do exist in the literature. One example is to start from a Lagrangian that contains both baryonic and quark degrees of freedom \cite{Marczenko:2020jma,Dexheimer:2020rlp}, another is a holographic approach where baryonic and quark phases are realized in a consistent way \cite{BitaghsirFadafan:2018uzs,Ishii:2019gta,Kovensky:2020xif}. Here we pursue a very simple idea, already put forward in Refs.\ \cite{Fraga:2018cvr,Schmitt:2020tac}: we start from a Lagrangian with only baryonic degrees of freedom, where the masses are entirely generated through the chiral condensate, similar to the extended linear sigma model employed in Refs.\ \cite{Heinz:2013hza, Haber:2014ula}. This allows us to observe a chiral phase transition and a (approximately) chirally symmetric phase at high densities with very small baryonic masses. This is in contrast to similar models of the Walecka type \cite{Walecka:1974qa,Boguta:1977xi,Sugahara:1993wz,Schaffner:1995th,Alford:2020pld}, which can only be used to describe chirally broken matter. Our study extends the model of Refs.\ \cite{Drews:2013hha,Drews:2014spa,Fraga:2018cvr,Schmitt:2020tac}
to include strangeness via hyperonic degrees of freedom, which gives rise to a more realistic picture of the chirally restored phase, resembling ``strange quark matter" in various aspects that will be discussed in detail. In particular for neutron star conditions this is an essential improvement since without strangeness the model does not have any degrees of freedom that carry both baryon number and negative electric charge. This is relevant due to the neutrality constraint and  can also be expected to alter the screening effects at the interfaces of mixed phases, and thus our study provides a framework to improve the study of ``chiral pasta" \cite{Schmitt:2020tac}. 

By including hyperons we do not necessarily change the baryonic phase of the model. Whether actual hyperons appear is decided dynamically.  They may be disfavored before the chiral phase transition, and we shall see that they indeed only appear for values of the model parameters  that are in conflict with astrophysical data of compact stars. However, the hyperonic degrees of freedom do play a role in the chirally restored phase and we shall see that parameter regions allowed by empirical constraints do also allow for strangeness in the chirally restored phase for all chemical potentials above the chiral phase transition. It is in this sense that we speak of strange quark matter from a baryonic approach, having in mind that there are no quark degrees of freedom in our model and that we should not expect to reproduce all known properties of weakly interacting, three-flavor quark matter at asymptotically large densities. Instead, our model provides a prediction for chirally restored matter close to the chiral phase transition, relevant for compact stars, with properties very different from simple extrapolations of weakly-coupled quark matter.  

We shall keep most of the approximations used in the non-strange version of the model \cite{Fraga:2018cvr,Schmitt:2020tac}, i.e. our evaluation will be in the mean-field and no-sea approximations at zero temperature, and we shall neglect Cooper pairing that is expected to occur in nuclear matter \cite{Sedrakian:2018ydt} and quark matter \cite{Alford:2007xm}. As in Refs.\ \cite{Fraga:2018cvr,Schmitt:2020tac}, we should keep in mind that our description of dense matter is based on extrapolating a model constructed mainly to reproduce low-density properties of nuclear matter.
We shall restrict ourselves to thermodynamic properties and homogeneous phases, within the constraints of equilibrium with respect to the weak interactions and local electric charge neutrality. The main idea of the paper is to set up the model and explore its parameter space in order to identify regions in which it reproduces basic properties of symmetric nuclear matter at saturation, basic properties of strange quark matter at asymptotically large densities and is able to reproduce compact stars with a mass of at least about 2.1 solar masses, meeting the constraint set by the heaviest known compact star \cite{NANOGrav:2019jur,Fonseca:2021wxt}. In doing so, we can e.g. constrain to a very narrow range the poorly known slope parameter of the symmetry energy, $L\simeq (88-92)\, {\rm MeV}$. Our study thus lays the ground for future studies for instance of the quark-hadron mixed phase or the chiral density wave \cite{Heinz:2013hza,Moreira:2013ura,Carignano:2019ivp} in the vicinity of the chiral phase transition. 

Our paper is organized as follows. We set up the model in Sec.\ \ref{sec:setup}, including the underlying Lagrangian and the resulting Euler-Lagrange equations. Some guidance and insight for the setup is gained from an $SU(3)$ symmetric approach, which we review in Appendix \ref{app:su3}. In Sec.\ \ref{sec:para} we discuss carefully the matching procedure of our parameters and identify the freedom in the parameter choices left by experimental uncertainties, mainly in the strangeness sector. Our main results are presented and discussed in Sec.\ \ref{sec:results}, which we have divided into a subsection on a few selected parameters sets, Sec.\ \ref{sec:4sets}, and a more general survey of the parameter space, Sec.\ \ref{sec:paraspace}, where we draw some parameter-independent conclusions. We give a summary and an outlook in Sec.\ \ref{sec:summary}.

\section{Setup}
\label{sec:setup}

\subsection{Lagrangian}

The hadronic part of our Lagrangian is composed of baryonic and mesonic contributions and baryon-meson interactions,  
\be
{\cal L} = {\cal L}_B + {\cal L}_M + {\cal L}_I  \, .
\ee
The baryonic part is 
\be
{\cal L}_B = \sum_i\bar{\psi}_i(i\gamma^\mu\partial_\mu+\gamma^0\mu_i)\psi_i \, , 
\ee
where $\bar{\psi}_i = \psi^\dag_i\gamma^0$ and the sum is over the baryon octet, $i=n,p,\Sigma^0,\Sigma^-,\Sigma^+,\Lambda,\Xi^0,\Xi^-$.  We have not included any explicit mass terms, all baryon masses will be generated dynamically by the chiral condensate. Since in QCD chiral symmetry is only approximate, adding small explicit masses does not violate general principles, and this was indeed done in comparable approaches  \cite{Dexheimer:2008ax}. For simplicity, and to avoid 
additional parameters, we shall account for explicit chiral symmetry breaking only in the meson potential and the choice of the baryon-meson coupling constants. The Lagrangian formally contains a chemical potential for each of the 8 baryon species, but in (three-flavor) QCD there are only three independent chemical potentials, associated with baryon number, isospin, and strangeness.  In terms of these chemical potentials, 
\be
\mu_i = \mu_B + I_i\mu_I + S_i\mu_S \, ,
\ee
where $I_i$ is the third component of the isospin and $S_i$ is the strangeness of the baryons, such that explicitly
\begin{subequations}
\bea
\mu_{n/p} &=& \mu_B\pm \mu_I \, , \\[2ex]
\mu_{\Sigma^\pm} &=& \mu_B\mp2\mu_I-\mu_S \, , \\[2ex]
\mu_\Lambda=\mu_{\Sigma^0}&=&\mu_B-\mu_S \, , \\[2ex]
\mu_{\Xi^-/\Xi^0}&=&\mu_B\pm \mu_I-2\mu_S \, .
\eea
\end{subequations}
The number of independent chemical potentials is further reduced by the conditions of equilibrium with respect to the weak interactions and electric charge neutrality. We require the leptonic process $p+e\to n+\nu_e$ to be in equilibrium with the inverse reaction $n\to p+e+\bar{\nu}_e$, and the same for the non-leptonic processes $n+n\leftrightarrow p+\Sigma^-$. Other weak reactions involving hyperons exist but their equilibration does not yield independent conditions for our chemical potentials. We shall assume that neutrinos have mean free paths larger than the size of the system, such that we may set the neutrino chemical potential to zero. This is a good assumption for neutron stars unless the temperature is larger than about a few MeV, which is only the case in  the very early stages of their evolution and in merger processes. At zero temperature, weak equilibrium directly translates into simple conditions for the chemical potentials, $\mu_p+\mu_e=\mu_n$, and $2\mu_n=\mu_p+\mu_{\Sigma^-}$. As a result, we can express $\mu_B$, $\mu_I$, $\mu_S$ in terms of neutron and electron chemical potentials,
\be
\mu_B = \mu_n-\frac{\mu_e}{2} \, , \qquad \mu_I = \frac{\mu_e}{2} \, , \qquad \mu_S = -\frac{\mu_e}{2} \, .    \label{muBIS}
\ee
The mesonic part of the Lagrangian contains the scalar meson $\sigma$ and the vector mesons $\omega^\mu$, $\rho_0^\mu$, $\phi^\mu$,  
\bea \label{Lmes}
{\cal L}_M &=& \frac{1}{2}\partial_\mu\sigma\partial^\mu\sigma - U(\sigma) -\frac{1}{4}\omega_{\mu\nu}\omega^{\mu\nu}-\frac{1}{4}\phi_{\mu\nu}\phi^{\mu\nu}-\frac{1}{4}\rho_{\mu\nu}^0\rho^{\mu\nu}_0+\frac{m_\omega^2}{2}\omega_\mu\omega^\mu+\frac{m_\phi^2}{2}\phi_\mu\phi^\mu+\frac{m_\rho^2}{2}\rho_\mu^0\rho^\mu_0\non[2ex]
&& +\frac{d}{4}(\omega_\mu\omega^\mu+\rho_\mu^0\rho^\mu_0+\phi_\mu\phi^\mu)^2\, , 
\eea
where $\omega_{\mu\nu} = \partial_\mu\omega_\nu - \partial_\nu\omega_\mu$ and analogously for $\phi_{\mu\nu}$ and $\rho^0_{\mu\nu}$. This Lagrangian can be viewed as a subset of the Lagrangian containing the full scalar, pseudoscalar, and vector meson nonets \cite{Lenaghan:2000ey}, only keeping the fields that we assume to condense in the medium given by the baryons. This is justified by the mean-field approximation, where the fluctuations of the meson fields are neglected. For instance, the pseudoscalar nonet is completely omitted here because we assume none of these fields to condense. It is only indirectly used by fitting one of the parameters of the potential $U$ to the pion mass. Moreover, in the scalar sector, the fields corresponding to the 0 and 8 direction with regard to the commonly used generators of $U(3)$ are usually rotated to give a non-strange scalar field $\sigma$ and a strange field $\zeta$. This is explained more explicitly in Appendix \ref{app:su3}, where we briefly review the more systematic approach using the full mesonic and baryonic multiplets. Here, in the main part, we omit the $\zeta$ field (and condensate) for simplicity. This is comparable to the approximation used in Walecka-like models, where the excitations of the scalar fields (not their condensates) are fundamental degrees of freedom of the Lagrangian. In this case, the strangeness sector, i.e., the excitation of the $\zeta$, is sometimes omitted as well for phenomenological reasons \cite{Weissenborn:2011kb,Lopes:2021jpm}. The potential for the remaining scalar meson is chosen to be the same as in the two-flavor version of this model \cite{Drews:2013hha,Drews:2014spa,Fraga:2018cvr,Schmitt:2020tac},
\be \label{U}
U(\sigma) = \sum_{n=1}^4 \frac{a_n}{n!} \frac{(\sigma^2-f_\pi^2)^n}{2^n}-\epsilon(\sigma-f_\pi) \, , 
\ee
with parameters $a_1$, $a_2$, $a_3$, $a_4$, $\epsilon$ and the pion decay constant $f_\pi \simeq 92.4\, {\rm MeV}$.  Temporarily including pion fluctuations, we fit $a_1=m_\pi^2$ to reproduce the vacuum mass of the pion $m_\pi=139\, {\rm MeV}$, and requiring the vacuum value of the chiral condensate to be $\langle\sigma\rangle=f_\pi$, we obtain $\epsilon=m_\pi^2f_\pi$. For the vector meson masses in Eq.\ (\ref{Lmes}) we will use $m_\omega=782\, {\rm MeV}$, $m_\phi=1020\, {\rm MeV}$, $m_\rho=775\, {\rm MeV}$. We have included a quartic meson coupling term \cite{Bodmer:1991hz,Horowitz:2002mb,Dexheimer:2018dhb,Dexheimer:2020rlp} with coupling constant $d\ge 0$, which 
 will play an important role for our results. The structure of this term 
 is a particular choice within the more general quartic term based on a chiral approach, see appendix \ref{app:su3} and in particular Eq.\ (\ref{d1d2}).

The baryon-meson interactions are given by
\bea
{\cal L}_I &=& -\sum_i \bar{\psi}_i(g_{i\sigma} \sigma + g_{i\omega}\gamma^\mu \omega_\mu +g_{i\rho}\gamma^\mu\rho^0_\mu +g_{i\phi}\gamma^\mu\phi_\mu )\psi_i
\,  .
\eea
As dictated by the chiral $SU(3)$ approach, the coupling constants within each isospin multiplet are related, see appendix \ref{app:su3}, and will be denoted by 
\bea
g_{Nx} \equiv g_{nx} = g_{px} \, , \qquad g_{\Sigma x}\equiv g_{\Sigma^0x} = g_{\Sigma^\pm x} \, , \qquad g_{\Xi x} \equiv g_{\Xi^0x} = g_{\Xi^-x} \, , 
\eea
for $x=\sigma, \omega, \phi$, and 
\bea
g_{N\rho} \equiv g_{n\rho} = -g_{p\rho} \, , \qquad g_{\Sigma \rho}\equiv  g_{\Sigma^+ \rho} =-g_{\Sigma^- \rho} \, , \qquad g_{\Xi \rho} \equiv g_{\Xi^0\rho} = -g_{\Xi^-\rho} \, , 
\eea
while $g_{\Sigma^0\rho} = g_{\Lambda\rho} = 0$. 
The coupling constants $g_{i\sigma}$ between the baryons and the scalar field are fixed by their vacuum masses. At mean-field level, and using that in the vacuum  $\langle\sigma\rangle=f_\pi$, the baryonic vacuum masses are $m_{i} = g_{i\sigma}f_\pi$. Using $m_N\equiv m_{n/p} \simeq 939\, {\rm MeV}$, $m_{\Lambda} \simeq 1115\, {\rm MeV}$, $m_{\Sigma^\pm/\Sigma^0}  \simeq 1190\, {\rm MeV}$,  $m_{\Xi^-/\Xi^0} \simeq 1315\, {\rm MeV}$, 
this fixes the coupling constants $g_{i\sigma}$. Fixing the 
couplings between the baryons and the vector mesons is more complicated. It is possible to derive the coupling terms from a $SU(3)$ invariant approach, see appendix \ref{app:su3}. We shall use the resulting constraints for some of the hyperonic couplings, combined with a phenomenological approach for the nucleonic couplings, as we shall explain in Sec.\ \ref{sec:para}.

\subsection{Free energy and stationarity equations}

We allow the scalar meson field and the temporal components of the vector meson fields to condense and denote the corresponding condensates by 
\be
\sigma\equiv \langle\sigma\rangle \, , \quad 
\omega\equiv \langle\omega_0\rangle \, , \quad 
\rho\equiv \langle\rho_0^0\rangle \, ,  \quad 
\phi\equiv \langle\phi_0\rangle \, .
\ee
They are assumed to be homogeneous in space, and we neglect all mesonic  fluctuations. This allows us to write down an effective ``mean-field Lagrangian",
\be
{\cal L}= \sum_i\bar{\psi}_i(i\gamma^\mu\partial_\mu+\gamma^0\mu_i^* - M_i)\psi_i- U(\sigma)-V(\omega,\rho,\phi) \, , 
\ee 
with the vector meson potential
\be
V(\omega,\rho,\phi)=-\frac{1}{2}(m_\omega^2\omega^2+m_\rho^2\rho^2+m_\phi^2\phi^2
) -\frac{d}{4}(\omega^2+\rho^2+\phi^2)^2 \, , 
\ee
the effective chemical potentials
\begin{subequations}\label{mustar}
\bea
\mu_{n/p}^*&=&\mu_{n/p}-g_{N\omega}\omega-g_{N\phi}\phi\mp g_{N\rho}\rho \, , \label{muNs}\\[2ex]
\mu_{\Sigma^0}^*&=&\mu_{\Sigma^0}-g_{\Sigma\omega}\omega-g_{\Sigma\phi}\phi \, , \\[2ex]
\mu_{\Sigma^\pm}^*&=&\mu_{\Sigma^\pm}-g_{\Sigma\omega}\omega-g_{\Sigma\phi}\phi\mp g_{\Sigma\rho}\rho \, , \\[2ex]
\mu_{\Lambda}^*&=&\mu_{\Lambda}-g_{\Lambda\omega}\omega-g_{\Lambda\phi}\phi \, , \\[2ex]
\mu_{\Xi^0/\Xi^-}^*&=&\mu_{\Xi^0/\Xi^-}-g_{\Xi\omega}\omega-g_{\Xi\phi}\phi\mp g_{\Xi\rho}\rho \, , 
\eea
\end{subequations}
and the effective, medium-dependent  masses
\bea
M_{n/p}&=&g_{N\sigma}\sigma\, , \qquad 
M_{\Sigma^0/\Sigma^\pm}=g_{\Sigma\sigma}\sigma\, , \qquad 
M_{\Lambda}=g_{\Lambda\sigma}\sigma\, , \qquad 
M_{\Xi^0/\Xi^-}=g_{\Xi\sigma}\sigma\, .\label{masses}
\eea
As often done in comparable phenomenological models, we shall omit the (renormalized) vacuum contribution (``no-sea approximation"). The idea is that this contribution would only yield a quantitative change and since the entire approach is of phenomenological nature there is not much to be gained from the inclusion of this contribution, given that the parameters of the model will be fitted  within this approximation to low-energy nuclear matter properties. (There are cases, however, where the vacuum part makes a {\it qualitative} difference, for instance in the case of a background magnetic field \cite{Fraga:2008qn,Mizher:2010zb,Endrodi:2013cs,Haber:2014ula}.) We shall also restrict ourselves to zero temperature throughout the paper. Then, the free energy density
becomes 
\be \label{Omega}
\Omega = -\sum_ip(\mu_i^*,M_i)+U(\sigma)+ V(\omega,\rho,\phi) -p(\mu_e,m_e) - p(\mu_\mu,m_\mu) \, ,
\ee
where the pressure of each fermion species is given by the function
\be
p(\mu,M) = \frac{\Theta(\mu-m)}{8\pi^2}\left[\left(\frac{2}{3}k_F^3-m^2k_F\right)\mu+m^4\ln\frac{k_F+\mu}{m}\right] \, ,
\ee
with the Fermi momentum 
\be
k_F=\sqrt{\mu^2-m^2} \, . 
\ee
In Eq.\ (\ref{Omega}) we have added the leptonic contribution, with electron and muon chemical potentials $\mu_e$, $\mu_\mu$, and their masses $m_e=0.511\, {\rm MeV}$ and $m_\mu = 106\, {\rm MeV}$. Weak equilibrium requires $\mu_e=\mu_\mu$, for instance through the processes $e\to \mu +\bar{\nu}_\mu+\nu_e$ and $\mu\to e+\bar{\nu}_e+\nu_\mu$.  
We define the following general expressions for the scalar density and the fermionic number density, 
\begin{subequations}
\bea
n_{\rm sc}(\mu,m) &\equiv & -\frac{\partial p}{\partial m} =  \Theta(\mu-m)\frac{m}{2\pi^2}\left(k_F\mu-m^2\ln\frac{k_F+\mu}{m}\right) \, , \\[2ex]
n(\mu,m) &\equiv & \frac{\partial p}{\partial \mu} = \Theta(\mu-m)\frac{k_F^3}{3\pi^2} \, . \label{nmum}
\eea
\end{subequations}
Then, the Euler-Lagrange equations can be written as
\begin{subequations} \label{stats}
\bea
0&=&\frac{\partial \Omega}{\partial\sigma} = \frac{\partial U}{\partial \sigma} +\sum_i g_{i\sigma} n_{{\rm sc},i} \, , \label{Eqsig}\\[2ex]
0&=&\frac{\partial \Omega}{\partial\omega} = \frac{\partial V}{\partial \omega} +\sum_i g_{i\omega}n_i \, , \label{Eqw}\\[2ex]
0&=&\frac{\partial \Omega}{\partial\rho} = \frac{\partial V}{\partial \rho} +\sum_ig_{i\rho} n_i \, , \\[2ex]
0&=&\frac{\partial \Omega}{\partial\phi} = \frac{\partial V}{\partial \phi} +\sum_i g_{i\phi}n_i \, , \label{Eqphi}
\eea
\end{subequations}
where $n_{{\rm sc},i}\equiv n_{\rm sc}(\mu_i^*,M_i)$ and $n_i\equiv n(\mu_i^*,M_i)$.
Additionally, we need the constraint from local electric charge neutrality, which reads 
\be \label{neutral} 
0 = \frac{\partial\Omega}{\partial\mu_e} = -n_p-n_{\Sigma^+}+n_{\Sigma^-}+n_{\Xi^-}+n_e+n_\mu \, .
\ee
For the equation of state we shall need the energy density
\be\label{epsilon}
\epsilon = -P +\mu_e n_e+\mu_\mu n_\mu + \mu_Bn_B+\mu_S n_S+\mu_In_I = -P + \mu_n n_B \, ,
\ee
where $P=-\Omega$ is the pressure,  where, in the second step, we have used the chemical potentials (\ref{muBIS}) and the charge neutrality condition (\ref{neutral}), and where baryon, strangeness, and isospin number densities are 
\be
n_B = \sum _i n_i \, ,\qquad  n_S = \sum_i S_i n_i \, ,\qquad n_I = \sum_i I_i n_i \, .
\ee

\subsection{Speed of sound}
\label{sec:sound}

 We require that our model reproduces the speed of sound of asymptotically dense cold QCD, such that our chirally restored phase shares this property with realistic quark matter. At asymptotically large densities the speed of sound squared $c_s^2$ of QCD goes to the conformal limit $1/3$, since in this limit $\mu_B$ is much larger than the QCD scale and, due to asymptotic freedom, also much larger than the constituent quark masses. Therefore, the baryon density is that of a free gas of fermions, $n_B\propto \mu_B^3$, which yields $c_s^2=1/3$, independent of the proportionality constant, as can be easily checked from the definition 
\be \label{cssq}
c_s^2 = \frac{\partial P}{\partial \epsilon} = \frac{n_B}{\mu_B}\left(\frac{d n_B}{d \mu_B}\right)^{-1} 
\,  .
\ee
Here, the first expression is valid in general, i.e., also for nonzero temperatures, in which case the derivative with respect to $\epsilon$ is taken at fixed entropy per particle. The second expression is valid at zero temperature; see for instance appendix E of Ref.\ \cite{BitaghsirFadafan:2018uzs} for a derivation of the general expression in terms of derivatives with respect to the chemical potential and temperature. 

To discuss the speed of sound in our model, let us for illustrative purposes in this section only consider isospin-symmetric nuclear matter without strangeness, i.e. we ignore hyperons for now and the only nonzero meson condensates are $\sigma$ and $\omega$. Also ignoring neutrality and a possible lepton contribution, the only relevant equations are Eqs.\ (\ref{Eqsig}) and (\ref{Eqw}), which have to be solved for $\sigma$ and $\omega$ and which we write as
\begin{subequations} \label{statsymf}
\bea
0&=&f_1(\sigma,\omega,\mu_B) \equiv \frac{\partial U}{\partial\sigma}+2g_{N\sigma}n_{\rm sc}(\mu_B^*,M) \, , \label{statsymf1}\\[2ex] 
0&=&f_2(\sigma,\omega,\mu_B) \equiv
\omega(m_\omega^2+d\omega^2)-g_{N\omega} n_B 
  \, ,
\label{statsymf2}
\eea
\end{subequations}
with $M=g_{N\sigma}\sigma$, $\mu_B^*=\mu_B-g_{N\omega}\omega$. 
For the speed of sound we need the derivative 
\be
\frac{d n_B}{d \mu_B}=\frac{\partial n_B}{\partial \mu_B}+\frac{\partial n_B}{\partial \sigma}\frac{\partial \sigma}{\partial \mu_B}+\frac{\partial n_B}{\partial \omega}\frac{\partial \omega}{\partial \mu_B} \, .
\ee
The explicit derivatives of $n_B$ are easily obtained, but $\sigma$ and $\omega$ are only given implicitly by Eqs.\ (\ref{statsymf}) (there is no analytical solution even in this simplified scenario). We can, however, compute the relevant derivatives in terms of $\sigma$ and $\omega$ via
\be
\left(\frac{\partial\sigma}{\partial\mu_B},\frac{\partial\omega}{\partial\mu_B}\right)=
-\left(\frac{\partial f_1}{\partial\mu_B},\frac{\partial f_2}{\partial\mu_B}\right)\left(\begin{array}{cc} \displaystyle{\frac{\partial f_1}{\partial\sigma}} & \displaystyle{\frac{\partial f_2}{\partial\sigma}} \\[2ex] \displaystyle{\frac{\partial f_1}{\partial\omega}} & \displaystyle{\frac{\partial f_2}{\partial\omega}} \end{array}\right)^{-1} \, .
\ee
Inserting all this into the definition of the speed of sound yields after some algebra
\be \label{cssqsym}
c_s^2 = \frac{1}{3}\frac{k_F^2}{\mu_B\mu_B^*}\left[\frac{\displaystyle{\frac{2k_F\mu_B^*}{\pi^2} 
+\frac{3}{M}\frac{\partial U}{\partial M}-
\frac{\partial^2U}{\partial M^2}}}{\displaystyle{\frac{2k_F^3}{\pi^2\mu_B^*}
+\frac{3}{M}\frac{\partial U}{\partial M}-\frac{\partial^2U}{\partial M^2}}}+\frac{2g_{N\omega}^2k_F\mu_B^*}{\pi^2(m_\omega^2+3d\omega^2)}\right] \, ,
\ee
where $k_F^2=(\mu_B^*)^2-M^2$. This relatively compact expression is valid for all densities, but still requires solving equations (\ref{statsymf}) numerically for an explicit evaluation. In this section we are only interested in the asymptotic limit, which can be evaluated analytically. One observes that
taking the limit $\mu_B\to \infty$ does not commute with the
limit $d\to 0$. If we first send $d\to 0$, the solutions of Eq.\ (\ref{statsymf}) become for large $\mu_B$
\be \label{swd0}
d=0: \qquad \sigma \simeq \left(\frac{2g_{N\omega}^2\pi}{3m_\omega^2}\right)^{2/3}\frac{f_\pi m_\pi^2}{g_{N\sigma}^2\mu_B^{2/3}} \, , \qquad \omega \simeq \frac{\mu_B}{g_{N\omega}}-\left(\frac{3\pi^2m_\omega^2\mu_B}{2g_{N\omega}^5}\right)^{1/3} \, .
\ee
The subleading term in $\omega$ is needed to obtain the leading behavior for $\mu_B^*$. Since for $\mu_B\to\infty$ we have $k_F\simeq \mu_B^*$, the first term in the square brackets in Eq.\ (\ref{cssqsym}) approaches 1. It is therefore subleading and the asymptotic speed of sound is given by the second term in the square brackets. With the relations (\ref{swd0}) we find $c_s^2 =1$. Therefore, if the quartic self-interactions are switched off in the Lagrangian, $d=0$, the speed of sound approaches the speed of light at asymptotically large $\mu_B$. 

On the other hand, if we first take the limit $\mu_B\to \infty$ at nonzero $d$ we find for the leading terms of the solution of Eqs.\ (\ref{statsymf})
\be \label{sigom2}
\sigma \simeq \left[1+\left(\frac{2g_{N\omega}^4}{3\pi^2 d}\right)^{1/3}\right]^2 \frac{f_\pi m_\pi^2\pi^2}{g_{N\sigma}\mu_B^2}  \, , \qquad \omega \simeq \left[1+\left(\frac{3\pi^2d}{2g_{N\omega}^4}\right)^{1/3}\right]^{-1} \frac{\mu_B}{g_{N\omega}}  \, .
\ee
Again, the first term in the square brackets in Eq.\ (\ref{cssqsym}) becomes 1, but this time it is of the same order as the second term, and both terms together give the asymptotic result $c_s^2=1/3$ for all $d>0$, a conclusion also reached for a similar model in Ref.\ \cite{Mueller:1996pm}. This shows that only in the presence of a quartic vector meson self-coupling our model reproduces the asymptotic speed of sound of QCD. 

These observations
also suggest that by choosing a sufficiently small but nonzero $d$, the speed of sound becomes arbitrarily close to 1 at intermediate densities. 
The reason is that the behavior of the condensates (\ref{swd0}) also holds in a regime where $\mu_B$ is large compared to all other energy scales while the dimensionless parameter $d\mu_B^2/m_\omega^2$ is small,
see Eqs.\ (\ref{statsymf2}) and (\ref{cssqsym}). For any nonzero $d$, of course, the behavior (\ref{sigom2}) eventually takes over as $\mu_B$ is increased and the speed of sound approaches 1/3 asymptotically.
This can be confirmed numerically, as well as the fact that these asymptotic limits derived here remain valid in the more complicated scenario including strangeness and the neutrality constraint.

\section{Parameter choices}
\label{sec:para}

Our strategy for fixing the parameters of the Lagrangian is to fit as many as possible to empirical vacuum and low-density quantities, and explore the parameter space of the remaining ones to understand the qualitative behavior of the model, in particular with respect to the chiral phase transition and the onset of strangeness. We have already used vacuum properties to fix $\epsilon, a_1, m_\omega, m_\phi, m_\rho, g_{N\sigma}, g_{\Lambda\sigma}, g_{\Sigma\sigma}, g_{\Xi\sigma}$. We assume that the nucleons do not couple to the hidden strangeness meson, $g_{N\phi}=0$ \cite{Papazoglou:1997uw,Dexheimer:2008ax,Sedrakian:2021goo}. It remains to choose values for, first,  $a_2,a_3,a_4,d,g_{N\omega},g_{N\rho}$, and, second, the couplings of the hyperons to the vector mesons $g_{\Lambda\omega},g_{\Sigma\omega},g_{\Xi\omega},
g_{\Lambda\phi},g_{\Sigma\phi},g_{\Xi\phi},g_{\Sigma\rho},g_{\Xi\rho}$. Let us discuss these two groups of parameters separately.

\subsection{Saturation properties}

We relate the 6 parameters $a_2,a_3,a_4,d,g_{N\omega},g_{N\rho}$ to  6 properties of isospin-symmetric nuclear matter at saturation: we use the well-known 
binding energy $E_B=-16.3\, {\rm MeV}$ and saturation density $n_0=0.153\, {\rm fm}^{-3}$, and also work with a definite symmetry energy $S=32\, {\rm MeV}$, following the empirical estimates $S\simeq (30.2-33.7)\, {\rm MeV}$ \cite{Danielewicz:2013upa,Lattimer:2014sga} (see, however, Ref.\ \cite{Reed:2021nqk}, which predicts a somewhat larger value based on measurements of the neutron skin thickness by the PREX collaboration \cite{PREX:2021umo}). The incompressibility at saturation is less well known, $K\simeq (200-300)\, {\rm MeV}$. In our main results we shall employ the value $K=250\, {\rm MeV}$. We have checked that our results do not change much under variations of $K$ in the empirically allowed range. There is much more sensitivity to the effective nucleon mass at saturation, $M_0$, and the slope $L$ of the symmetry energy with respect to density changes away from saturation. For later, we shall keep in mind an empirical range of $M_0\simeq (0.7-0.8)m_N$ \cite{1989NuPhA.493..521G,Johnson:1987zza,Li:1992zza,glendenningbook,Jaminon:1989wj,Furnstahl:1997tk}. Estimates for the slope of the symmetry energy range from $L\simeq (40-60)\, {\rm MeV}$ \cite{Lattimer:2012xj,Oertel:2016bki,Tews:2016jhi} to more recent values using the result of the PREX experiment  \cite{PREX:2021umo}, indicating that larger values might be favored, $L\simeq (70-140)\, {\rm MeV}$  \cite{Reed:2021nqk}; for a recent overview of the various estimates for $L$ see Ref.\ \cite{Sotani:2022hhq}.

To set up the relation between the model parameters and the saturation properties, we denote the chemical potential at the onset of isospin-symmetric (non-strange) baryonic matter by $\mu_0=922.7\, {\rm MeV}$, and the effective baryon chemical potential by $\mu_0^* =\mu_n^*=\mu_p^* = \sqrt{k_F^2+M_0^2}$, where the Fermi momentum can be expressed in terms of the saturation density via $n_0=2k_F^3/(3\pi^2)$, which yields $k_F\simeq 260\, {\rm MeV}$.
In the absence of hyperons, baryon and isospin densities are  $n_B=n_n+n_p$ and $n_I=n_n-n_p$, respectively. In symmetric nuclear matter, where $n_I=0$, the stationarity equations (\ref{stats}) give $\rho=\phi=0$, while $\omega$ obeys the cubic equation 
\be \label{wcubic}
g_{N\omega}n_0=m_\omega^2\omega +d\omega^3\, , 
\ee
whose relevant solution we write as
\be \label{omegaf}
\omega_0= \frac{g_{N\omega}n_0}{m_\omega^2}f(x_0)\, , 
\ee
with 
\be
f(x) \equiv \frac{3}{2x}\frac{1-(\sqrt{1+x^2}-x)^{2/3}}{(\sqrt{1+x^2}-x)^{1/3}} \, , \qquad x_0\equiv \frac{3\sqrt{3d}\,g_{N\omega}n_0}{2m_\omega^3} \, . 
\ee
With $\lim_{x\to 0} f(x)=1$ we recover the case without quartic vector meson interactions, $d=0$. We also need the definitions of incompressibility, symmetry energy, and slope of the symmetry energy,
\be
K = 9n_B\frac{\partial\mu_B}{\partial n_B} \, , \qquad S =  \frac{n_B}{2} \frac{\partial\mu_I}{\partial n_I} \, , \qquad L = 3n_B\frac{\partial S}{\partial n_B} \, , \label{KSL}
\ee
where $K$ is evaluated for symmetric nuclear matter, the derivative in $S$ is taken at fixed $n_B$ and evaluated at $n_I=0$, and the derivative in $L$ is taken at fixed $n_I=0$. 

Putting all of this together, we obtain the following six conditions for the model parameters: 
\begin{subequations}\label{paramatch}\allowdisplaybreaks
\bea
g_{N\omega}^2 &=& \frac{m_\omega^2}{2n_0}(\mu_0-\mu_0^*)\left[1+\sqrt{1+\frac{4dn_0(\mu_0-\mu_0^*)}{m_\omega^4}}\right] \, ,\label{gNw} \\[2ex]
g_{N\rho}^2 &=& \frac{3\pi^2m_\rho^2}{k_F^3}\left(S-\frac{k_F^2}{6\mu_0^*}\right)\left(1+\frac{d\omega_0^2}{m_\rho^2}\right) \, ,\label{gNrho}\\[2ex]
L &=& \frac{3g_{N\rho}^2n_0}{2(m_\rho^2+d\omega_0^2)}\left[1-\frac{2d\,n_0g_{N\omega}\omega_0}{(m_\rho^2+d\omega_0^2)(m_\omega^2+3d\omega_0^2)}\right]+\frac{k_F^2}{3\mu_0^*}\left(1-\frac{K}{6\mu_0^*}\right)+\frac{g_{N\omega}^2n_0k_F^2}{2m_\omega^2\mu_0^{*2}}[f(x_0)+x_0f'(x_0)] \, ,\label{L}\\[2ex]
K &=&\frac{6k_F^3}{\pi^2}\frac{g_{N\omega}^2}{m_\omega^2}[f(x_0)+x_0f'(x_0)]
+\frac{3k_F^2}{\mu_0^*}-\frac{6k_F^3}{\pi^2}\left(\frac{M_0}{\mu_0^*}\right)^2
\left[\frac{1}{g_{N\sigma}^2}\frac{\partial^2U}{\partial \sigma^2}+\frac{2}{\pi^2}\int_0^{k_F} \frac{dk\,k^4}{(k^2+M_0^2)^{3/2}}\right]^{-1} \, ,\label{incomp} \\[2ex]
0&=& \frac{m_\omega^2}{2}\omega_0^2+\frac{d}{4}\omega_0^4-U(\sigma)+\frac{1}{4\pi^2}\left[\left(\frac{2}{3}k_F^3-M_0^2k_F\right)\mu_0^*+M_0^4\ln\frac{k_F+\mu_0^*}{M_0}\right] \, , \label{Psat}\\[2ex]
0&=&  \frac{\partial U}{\partial\sigma} + \frac{g_{N\sigma}M_0}{\pi^2}\left(k_F\mu_0^*-M_0^2\ln\frac{k_F+\mu_0^*}{M_0}\right) \, .\label{sigsat}
\eea
\end{subequations}
Here, the first relation is obtained from inserting the relation $\mu_0^*=\mu_0-g_{N\omega}\omega_0$, which follows from Eq.\ (\ref{muNs}), into Eq.\ (\ref{wcubic}); the next relations are obtained by computing $S$, $L$, and $K$ from their definitions (\ref{KSL}); finally, we have the condition that the pressure at saturation be identical to the pressure of the vacuum, which in our convention is zero, and the stationarity equation (\ref{Eqsig}) for $\sigma$, whose value is $\sigma=M_0/g_{N\sigma}$ at saturation.

For given $L,S,K,M_0,\mu_0,n_0$, Eqs.\ (\ref{paramatch}) can now be solved to obtain the model parameters $a_2,a_3,a_4,g_{N\omega},g_{N\rho},d$. For the practical calculation it is useful to note that (\ref{gNw}), (\ref{gNrho}), (\ref{L}) do not 
depend on $a_2$, $a_3$, $a_4$ (which only enter through the meson potential $U$), such that they can be solved separately for $g_{N\omega},g_{N\rho},d$. The results are then used to solve Eqs.\ (\ref{incomp}), (\ref{Psat}), (\ref{sigsat})  for $a_2$, $a_3$, $a_4$. 
If the quartic coupling is set to zero, $d=0$,  Eqs.\ (\ref{gNw}) and (\ref{gNrho}) can be used to obtain $g_{N\omega}$ and $g_{N\rho}$, and the coupled equations (\ref{incomp}), (\ref{Psat}), (\ref{sigsat}),  are used to fix $a_2,a_3,a_4$, while $L$ can only be computed afterwards, i.e., in this case there is no freedom in the parameter set to reproduce a given value for $L$.

\begin{figure} [t]
\begin{center}
\includegraphics[width=0.5\textwidth]{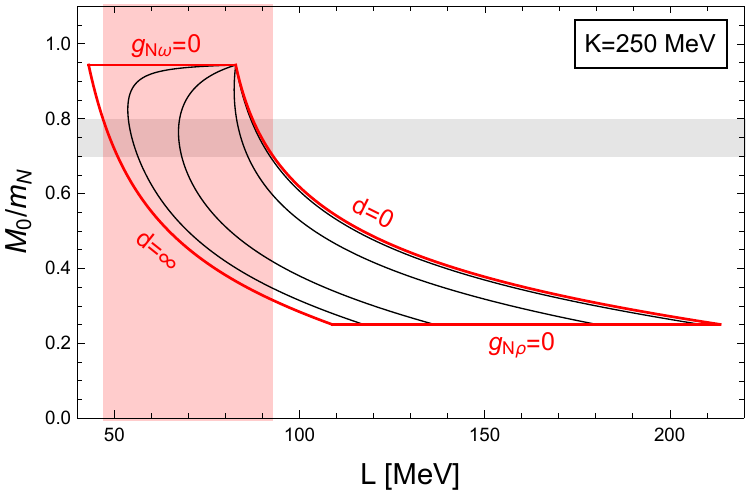}
\caption{Allowed window (red curves) of the model in the $M_0$-$L$ plane bounded by the limits (\ref{boundsM0}) and (\ref{L0inf}), solely derived from known saturation properties of symmetric nuclear matter. Here we have set $K=250\, {\rm MeV}$, which we will use throughout the paper. The thin black curves are lines of constant $d$, $d=10,10^2,10^3,10^4$ from right to left. The grey horizontal band indicates the empirically favored range for $M_0$, which results in a predicted range $L\simeq (47-93)\, {\rm MeV}$, shown by the red vertical band. }
\label{fig:M0Lspace}
\end{center}
\end{figure}

Interestingly, Eqs.\ (\ref{paramatch}) can be used to compute   
a window in the $M_0$-$L$ plane for a given value for $K$. From Eqs.\ (\ref{gNw}) and (\ref{gNrho}) we see that 
in order for $g_{N\omega}^2$ and $g_{N\rho}^2$ to be positive we need
\be \label{boundsM0}
k_F\sqrt{\left(\frac{k_F}{6S}\right)^2-1}\; < \;M_0 \;< \;\sqrt{\mu_0^2-k_F^2} \, .
\ee
We can also compute the limits of $L$ for $d=0$ and $d\to \infty$, which 
gives the range 
\be \label{L0inf}
S+\frac{k_F^2(3\mu_0-K)}{18\mu_0^{*2}} \;<\; L \;<\; \frac{3g_{N\rho}^2n_0}{2m_\rho^2}+\frac{k_F^2}{3\mu_0^*}\left(1-\frac{K}{6\mu_0^*}\right)+\frac{g_{N\omega}^2n_0k_F^2}{2m_\omega^2\mu_0^{*2}} \, ,
\ee
where the lower (upper) limit comes from $d\to\infty$ ($d=0$). We have considered the possibility of negative $d$, but have not found any physically sensible solutions, in most cases indicated by a superluminal speed of sound combined with the solutions of the stationarity equations turning complex at large densities, see also Refs.\ \cite{Mueller:1996pm,Alford:2022bpp}. 
The resulting window in the $M_0$-$L$ plane is shown in Fig.\ \ref{fig:M0Lspace} for $K=250\, {\rm MeV}$ (with all other saturation properties as given above). If we apply the realistic window $M_0\simeq (0.7-0.8)m_{N}$ we see that this already constrains the range for the slope of the symmetry energy to $L\simeq (47-93)\, {\rm MeV}$, as indicated by the shaded bands in the figure.

\subsection{Couplings between hyperons and vector mesons}
\label{sec:couplings}

The choice for the hyperon couplings $g_{\Lambda\omega},g_{\Sigma\omega},g_{\Xi\omega},
g_{\Lambda\phi},g_{\Sigma\phi},g_{\Xi\phi},g_{\Sigma\rho},g_{\Xi\rho}$ is much less constrained by experimental data. Here our strategy is to combine phenomenological constraints with the relations given by the chiral approach of appendix \ref{app:su3}, while leaving one degree of freedom to be varied to probe the dependence of our results on different choices of the hyperon couplings. The connection between the coupling constants and (potential) experimental data is made by the hyperon potential depths. The potential depth $U_i^{(j)}$ of a single hyperon $i$ in a medium of baryon species $j$ at arbitrary baryon density $n_B$ is computed as follows. We assume isospin-symmetric media, such that $n_p=n_n$ for $j=N$, $n_{\Sigma^+}=n_{\Sigma^0}=n_{\Sigma^-}$ for $j=\Sigma$, and $n_{\Xi^0}=n_{\Xi^-}$ for $j=\Xi$. As a consequence, $\rho=0$ in each case, and the Fermi momentum $k_F$ is related to the baryon density by
\be\label{nBkF}
n_B = \frac{s k_F^3}{3\pi^2} \, , 
\ee
where $s$ is a degeneracy factor, $s=2,1,3,2$ for baryonic media $N,\Lambda,\Sigma,\Xi$, respectively. The single-baryon energy $E_{k,i}^{(j)}$ of baryon $i$ in a medium of baryon $j$ obeys the relation
\be
E_{k,i}^{(j)} - \mu_i =\sqrt{k^2+(M_i^{(j)})^2}-\mu_i^{*(j)} \, ,
\ee
where $M_i^{(j)}$ is the medium-dependent mass of baryon $i$ and $\mu_i^{*(j)}$ is its effective chemical potential, containing the actual chemical potential $\mu_i$ and the medium-dependent condensates, see Eq.\ (\ref{mustar}). The potential is given by the minimum of the single-baryon energy $E_{k=0,i}^{(j)}$ minus the vacuum mass $m_i$, 
\be \label{Uij}
U_i^{(j)}=M_i^{(j)}-\mu_i^{*(j)}+\mu_i-m_{i} = g_{i\sigma}(\sigma^{(j)}-f_\pi)+g_{i\omega}\omega^{(j)}+g_{i\phi}\phi^{(j)} \, , 
\ee
where, in the second step, we have expressed the vacuum mass in terms of the vacuum value of the chiral condensate, $m_i=g_{i\sigma}f_\pi$. The 
medium-dependent mass and effective chemical potential have been written  in terms of the meson condensates in the medium of baryon $j$, which  
have to be computed numerically with the help of the stationarity equations at the given baryon density $n_B$ (\ref{nBkF}). 
For our purposes, Eq.\ (\ref{Uij}) is only needed for the hyperon potentials in a medium of nucleons at saturation density. In this case $\phi=0$, and using $g_{N\sigma}\sigma^{(N)}=M_0$, $g_{N\sigma}f_\pi = m_N$ we can write 
 \be \label{UiN}
U_i^{(N)} = \frac{g_{i\sigma}}{g_{N\sigma}}(M_0-m_N) + g_{i\omega}\omega_0\, , 
\ee
where $\omega^{(N)}=\omega_0$ is the value of the condensate at saturation (\ref{omegaf}). We thus have three relations, $i=\Sigma,\Lambda,\Xi$, to relate three hyperon potentials to the 
hyperon-omega coupling constants. 

In all our results we shall use the value 
\be \label{ULN}
U_\Lambda^{(N)} = -30\, {\rm MeV} \, , 
\ee
as suggested by experimental data \cite{PhysRevC.38.2700,Gal:2016boi} and adopted in comparable models \cite{PhysRevC.62.034311,Weissenborn:2011kb,Thapa:2020ohp}. The potentials for $\Sigma$ and $\Xi$ are less well known experimentally, with chiral effective theory suggesting $U_\Xi^{(N)}$ to have a relatively small absolute value with either sign possible and $U_\Sigma^{(N)}$ more likely to be positive \cite{Haidenbauer:2018gvg,Haidenbauer:2019boi}. For simplicity, we shall assume the values of both potentials to be identical,  
\be \label{calU}
{\cal U}\equiv U_\Sigma^{(N)} =  U_\Xi^{(N)} \, , 
\ee
 and vary ${\cal U}$ within a reasonable range. We shall see that within this simplistic approach we will have to choose in particular $U_\Sigma^{(N)}$ to be different from what is usually adopted. Due to the large uncertainties in our knowledge of these potentials this may not seem too unreasonable. Moreover, empirical constraints drive our choice to more attractive potentials compared to the most common values in the literature, such that one might expect hyperons to be unusually  favored in our results. However, we shall see that for the parameter sets that meet astrophysical constraints strangeness does not occur in the chirally broken phase. Therefore, even if the hyperon potentials we choose are different from their value in nature, we do not have hyperons with unphysical properties in our system. 
  The hyperon coupling constants then rather characterize the interactions in the chirally restored phase (i.e., of ``strange quark matter"), for which no direct experimental information is available and where astrophysical data are our best source for constraints, forcing us to somewhat stretch the usual regime for the hyperon potentials. 
 
After choosing a value of ${\cal U}$, Eqs.\  (\ref{UiN}), (\ref{ULN}), (\ref{calU}) fix the $\omega$ coupling constants $g_{\Sigma\omega}$, $g_{\Lambda\omega}$, $g_{\Xi\omega}$. This leaves the coupling constants
$g_{\Lambda\phi},g_{\Sigma\phi},g_{\Xi\phi},g_{\Sigma\rho},g_{\Xi\rho}$,  which we compute from the chiral relations (\ref{gchiral2}) (ignoring the relations in that equation for $g_{\Sigma\omega}$, $g_{\Lambda\omega}$, $g_{\Xi\omega}$).

\section{Results}
\label{sec:results}

We present and discuss our results as follows. First, in Sec.\ \ref{sec:4sets} we choose four parameter sets in order to demonstrate qualitatively different scenarios with respect to the chiral phase transition and the onset of strangeness that our model can produce. At this point, we do not yet discard parameter regions disfavored by astrophysical  data. The reason is that it is instructive to see that different scenarios can be realized in principle, keeping in mind that our model is of phenomenological nature. Therefore, a scenario realized in the present version of the model that appears to be excluded by data may be allowed in an improved version of the model, or in a different phenomenological model -- or in QCD. Then, second, in Sec.\ \ref{sec:paraspace}, we {\it do} discuss the empirical and astrophysical constraints systematically, which will lead to conclusions independent of the particular parameter choices.

\subsection{Selected parameter sets}
\label{sec:4sets}

\begin{table}[t]
\begin{tabular}{||c | c | c | c | c | c | c | c || c | c ||c||} 
 \hline
   $g_{N\omega}$ & $g_{N\rho}$ & $g_{\Lambda\omega}$ & $g_{\Sigma\omega}$ & $g_{\Xi\omega}$ & $a_2$ & $a_3 [{\rm MeV}^{-2}]$ & $a_4 [{\rm MeV}^{-4}]$ & $\;$$M_0/m_N$$\;$ & $\;$$L [{\rm MeV}]$$\;$ & $\;$Figs.\ \ref{fig:MNmun} -- \ref{fig:MR}$\;$ \\ [0.5ex] 
 \hline\hline
  $\;$ 10.23 $\;$ & $\;$4.138$\;$ & $\;$14.53$\;$ & $\;$14.59$\;$ & $\;$16.39$\;$ & $\;$44.69$\;$ & $\;$2.917$\cdot 10^{-4}$$\;$ & $\;$5.071$\cdot 10^{-5}$$\;$ &  0.72 & 89.91 & black\\ 
 \hline
  8.196 & 4.297 & 12.35 & 12.03 & 13.63 & 55.15 & -7.465$\cdot 10^{-3}$ & 9.553$\cdot 10^{-5}$ &0.8 & 86.24 &\textcolor{red}{red} \\
 \hline
  6.610 & 4.379 & 10.86 & 10.17 & 11.65 & 73.15 & 3.120$\cdot 10^{-2}$ & 2.865$\cdot 10^{-4}$ & 0.85 & 84.66 &\textcolor{blue}{blue}\\
 \hline
  3.291 & 4.477 & 9.371 & 7.155 & 8.738 & 465.5 & 3.643 & 1.305$\cdot 10^{-2}$ &0.92 & 83.14 &\textcolor{darkgreen}{green}\\
 \hline
\end{tabular}
\caption{Parameter sets used in Sec.\ \ref{sec:4sets}, Figs.\ \ref{fig:MNmun} -- \ref{fig:MR}. We have included the resulting values for the Dirac mass at saturation $M_0$ and the slope parameter $L$, 
while $K=250\, {\rm MeV}$ in all four cases. The parameters $\epsilon, a_1, m_\omega, m_\phi, m_\rho, g_{N\sigma}, g_{\Lambda\sigma}, g_{\Sigma\sigma},  g_{\Xi\sigma}$ are the same in all cases and fixed by vacuum properties as explained in Sec.\ \ref{sec:setup}. Moreover, in all four cases $d=21$, and the hyperon couplings listed here are chosen to give ${\cal U}=-50\, {\rm  MeV}$. The remaining hyperon-meson couplings $g_{\Lambda\phi},g_{\Sigma\phi},g_{\Xi\phi},g_{\Sigma\rho},g_{\Xi\rho}$ are  determined by the chiral relations (\ref{gchiral2}) in each case separately. }
\label{table:para}
\end{table}

We start with the four parameter sets specified in Table \ref{table:para}. They all give a potential ${\cal U} = -50\, {\rm MeV}$ for the $\Sigma$ and $\Xi$, and the quartic meson self-coupling constant is fixed to $d=21$. As mentioned above, we also keep the incompressibility at saturation fixed to $K=250\, {\rm MeV}$. The parameter sets are then obtained by varying the Dirac mass at saturation from low (approximately the lower end of the empirically allowed range) to high (somewhat larger than the empirically allowed maximum). The slope parameter $L$ then adjusts accordingly (varying, however, only by a few percent for the given choices). Note that $a_4$ turns out to be positive in all four cases as it should be since this ensures a bounded vacuum potential for $\sigma$.

\subsubsection{Chiral transition and onset of strangeness}
\label{sec:thermo}

\begin{figure} 
\begin{center}
\includegraphics[width=\textwidth]{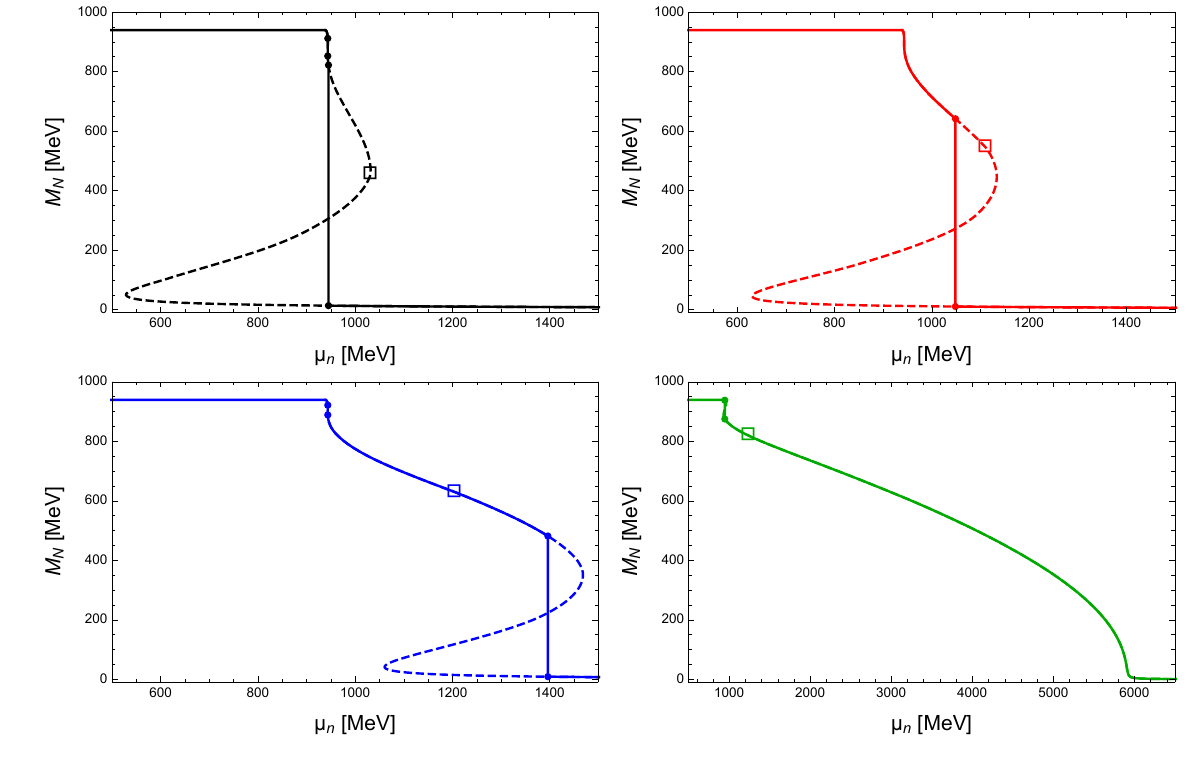}

\caption{Effective nucleon mass as a function of the neutron chemical potential for the four parameter sets given in Table \ref{table:para}. Solid lines correspond to stable phases, while the dashed segments are metastable ($M_N$ decreasing with $\mu_n$) or unstable ($M_N$ increasing with $\mu_n$). The open squares mark the onset of strangeness, and the dots mark the phase transition within the baryonic phase (upper left and both lower panels) and the chiral phase transition (both upper panels and lower left). In the bottom right panel the chiral transition has become a (steep) crossover.}
\label{fig:MNmun}
\end{center}
\end{figure}

\begin{figure} 
\begin{center}
\includegraphics[width=0.49\textwidth]{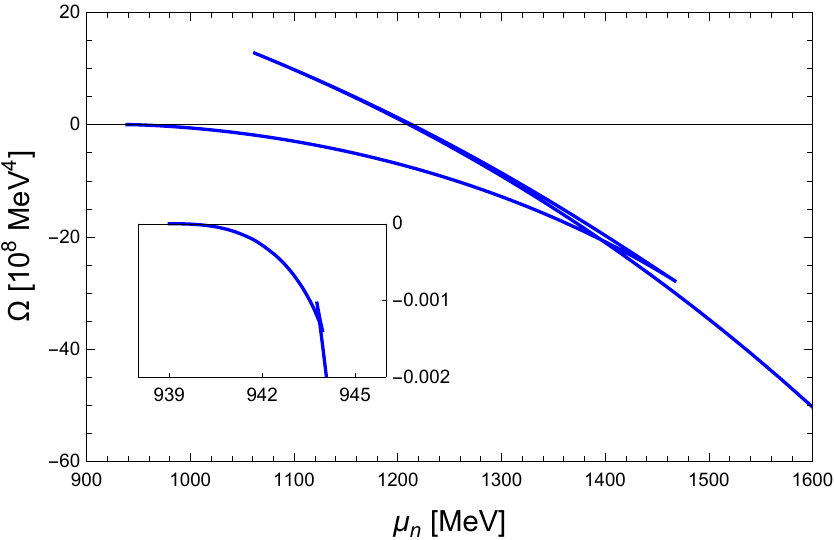}
\caption{Free energy density as a function of the neutron chemical potential for the third case (blue) of Fig.\ \ref{fig:MNmun}. The large three-valued region is the spinodal region of the first-order chiral phase transition, while the zoom-in shows a (weak) first-order transition within the chirally broken phase. }
\label{fig:Om}
\end{center}
\end{figure}

In Fig.\ \ref{fig:MNmun} we show the effective nucleon mass $M_N\equiv M_{n/p}$ as a function of the neutron chemical potential, obtained by solving the stationarity equations (\ref{stats}) together with the neutrality constraint (\ref{neutral}) numerically for $\sigma$, $\omega$, $\phi$, $\rho$, $\mu_e$ at given $\mu_n$ (and $T=0$). Since all baryon masses are proportional to the chiral condensate $\sigma$ (multiplied by a coupling constant to reproduce the vacuum masses), the effective hyperon masses follow the same behavior. The figure shows all branches of the solution, including the unstable and metastable ones. In all cases, there is an approximately chirally symmetric phase at large chemical potentials, where the baryon masses are very small. In three of the four cases shown here, the chirally restored phase is reached via a first-order phase transition. The location of the phase transition has to be determined from the free energy, i.e., by inserting the solutions of the stationarity equations back into the free energy density (\ref{Omega}). An example is shown in Fig.\ \ref{fig:Om}, corresponding to the lower left panel in Fig.\ \ref{fig:MNmun}. Determining the state with the lowest free energy at each $\mu_n$  allows us to identify the stable branches, shown as solid curves in Fig.\ \ref{fig:MNmun}. 

Besides the very prominent chiral phase transition, Fig.\ \ref{fig:Om} also shows a much weaker first-order phase transition at relatively low densities within the chirally broken phase. This phase transition can be understood as a ``remnant" of the first-order onset of isospin-symmetric nuclear matter. In that case, the free energy is multi-valued at the onset, and moving towards more neutron-rich matter tends to diminish this  multivaluedness, i.e., decrease the spinodal region. This happens gradually, and thus, even in the neutron-rich environment obtained here by the conditions of weak equilibrium and charge neutrality, it is possible that the spinodal region survives. This is the case in three of the four cases in Fig.\ \ref{fig:MNmun}, as indicated by the dots that mark the effective nucleon mass on either side of the transition. In contrast to the chiral transition, the curves of stable and unstable phases in the vicinity of this transition are not distinguishable by naked eye on the given scale.

 Fig.\ \ref{fig:MNmun} also indicates the onset of strangeness (open squares). We see that there are qualitatively different cases with respect to that onset (and demonstrating these differences is one main motivation for our choice of parameter sets): in the two upper panels, the onset of strangeness occurs in the metastable or unstable regime. This implies that the baryonic phase does not contain any hyperons, while strangeness appears immediately after the chiral transition, i.e. the transition is from nuclear matter to ``strange quark matter". Showing the possibility of this scenario within a model based on baryonic degrees of freedom has been one of the main goals of this paper (and we shall see below that astrophysical constraints favor this case). The precise location of the strangeness onset within the metastable/unstable regime is irrelevant for the stable, homogeneous phases discussed in this paper. However, it would be interesting for future studies to see how this location affects the properties of inhomogeneous phases, such as a mixed phase, which does know about the behavior of the model away from the stable branches. In the lower left panel of Fig.\ \ref{fig:MNmun}, strangeness occurs already in the baryonic phase. Therefore, in this case the sequence of phases is nuclear matter $\to$ hyperonic matter $\to$ chirally restored matter with strangeness. 
Finally, the lower right panel shows yet another qualitatively different behavior, namely a chiral crossover. In this case, strangeness occurs deeply in the baryonic regime (judging from the effective nucleon mass, which is about $800\, {\rm MeV}$ at that point). Then, there is a continuous transition to the phase with light degrees of freedom. It is striking that, first, this transition is still relatively ``sharp". It is difficult to distinguish it by naked eye  from a weak first-order transition. And, second, this sharp transition occurs at extremely large chemical potentials, much larger than in the interior of neutron stars. We have not found any parameter set with reasonable low-density properties that shows a significantly smoother crossover or a significantly smaller transition density. (Judging from the results of the non-strange, isospin-symmetric version of our model \cite{Fraga:2018cvr}, a much larger incompressibility, far beyond the physical range, is needed for such a scenario.) Nevertheless, it is interesting that our model allows for the possibility of a crossover, which is conceivable within QCD and corresponding model equations of state have been constructed \cite{Baym:2017whm,Baym:2019iky}, although this question becomes more subtle in the presence of Cooper pairing \cite{Schafer:1998ef,Hatsuda:2006ps,Schmitt:2010pf,Cherman:2018jir}.

\begin{figure} 
\begin{center}
\includegraphics[width=\textwidth]{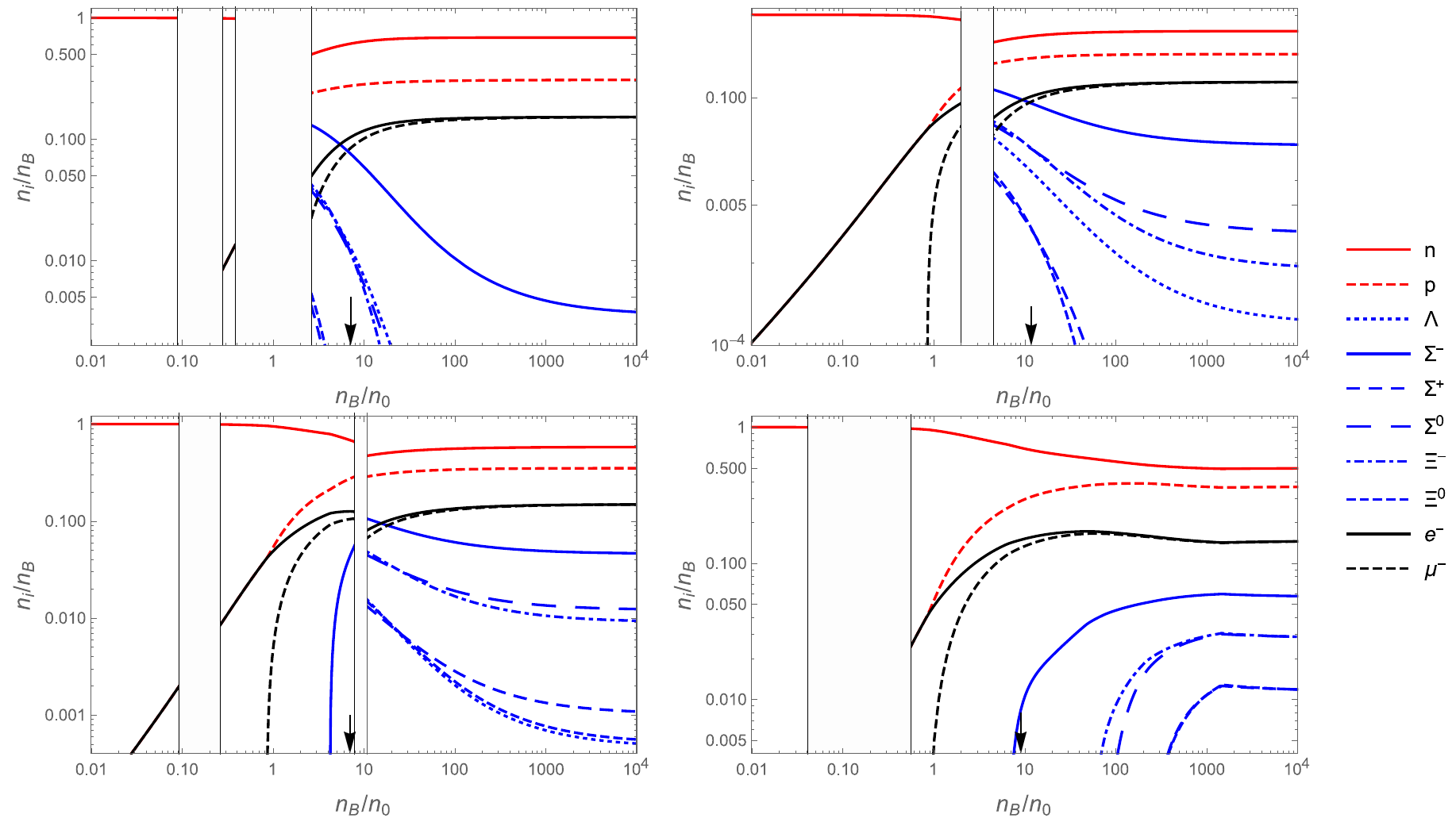}
\caption{Density fractions as a function of baryon density (normalized to the saturation density of symmetric nuclear matter $n_0$) for the four cases of Fig.\ \ref{fig:MNmun}, showing non-strange baryons (red), strange baryons (blue) and leptons (black). First-order transitions appear in the form of a gap in the horizontal direction since the density is discontinuous. Cusps arise from the onset of baryonic species. The arrow on the horizontal axis marks the central density of the most massive star possible for each parameter set. }
\label{fig:fractions}
\end{center}
\end{figure}

While the onset of strangeness marked in Fig.\ \ref{fig:MNmun} refers to the first strange degree of freedom, Fig.\ \ref{fig:fractions} shows all individual particle fractions as functions of density. We have distinguished non-strange baryons from hyperons and leptons by the color of the curves to facilitate the interpretation. Since the horizontal axis represents density, there are disallowed regions due to the first-order phase transitions. There are metastable and unstable branches in these regions which we have omitted since they are not very instructive. Also, the disallowed regions can be populated by inhomogeneous mixed phases, which we are ignoring in this paper. We see that the lower critical density of the chiral phase transition varies greatly between the different parameter sets, occurring as early as $n_B\simeq 0.4\, n_0$ in the upper left panel. We have also marked the maximal central densities reached in compact stars for each case by an arrow. These densities lie somewhere in the range $n_B\sim (7-10)\, n_0$, a somewhat large number compared to most comparable phenomenological models.

The figure also shows that the most prevalent strange degree of freedom in all four cases is the $\Sigma^-$, which is the lightest non-leptonic degree of freedom with negative electric charge. We also see that in the cases with a first-order chiral phase transition the density fractions of the strange degrees of freedom decrease as the density is increased. This is perhaps somewhat unexpected, at least having in mind the following simple picture of quark matter: At intermediate densities we expect the constituent mass of the strange quark to be larger than that of the up and down quarks. At ultra-high densities, due to asymptotic freedom, the quark masses approach the current mass limit, whose scale becomes negligible compared to the chemical potential. As a consequence, one might expect the strangeness content to {\it increase} as one moves to higher densities, although the strong-coupling nature of the problem at intermediate densities does not allow a firm first-principles prediction for this behavior. What {\it is} firmly predicted by QCD, however, is that three-flavor quark matter becomes flavor symmetric at asymptotically large densities. Our results in Fig.\ \ref{fig:fractions} show two interesting properties of asymptotically dense matter. First, a nonzero amount of strangeness survives asymptotically. The parameter sets are chosen deliberately to ensure this property, and we shall discuss in the subsequent section that this is not the case for all parameter choices. Second, our asymptotic matter is clearly not flavor symmetric, i.e., the up, down, and strange content of our baryonic degrees of freedom is not equal. We show in Appendix \ref{app:symmetry} that there {\it are} choices of the hyperon-meson coupling constants that lead to asymptotic flavor symmetry (while keeping the saturation properties of symmetric nuclear matter fixed). This would be desirable in our context since this would make our chirally restored matter even more similar to actual QCD quark matter. However, we have not found parameter sets that at the same time produce sufficiently heavy neutron stars, and thus here, in the main part, we do not work with the parameter constraints derived in Appendix \ref{app:symmetry}.

\subsubsection{Speed of sound and mass-radius curves}
\label{sec:stars}

\begin{figure} [t]
\begin{center}
\includegraphics[width=0.49\textwidth]{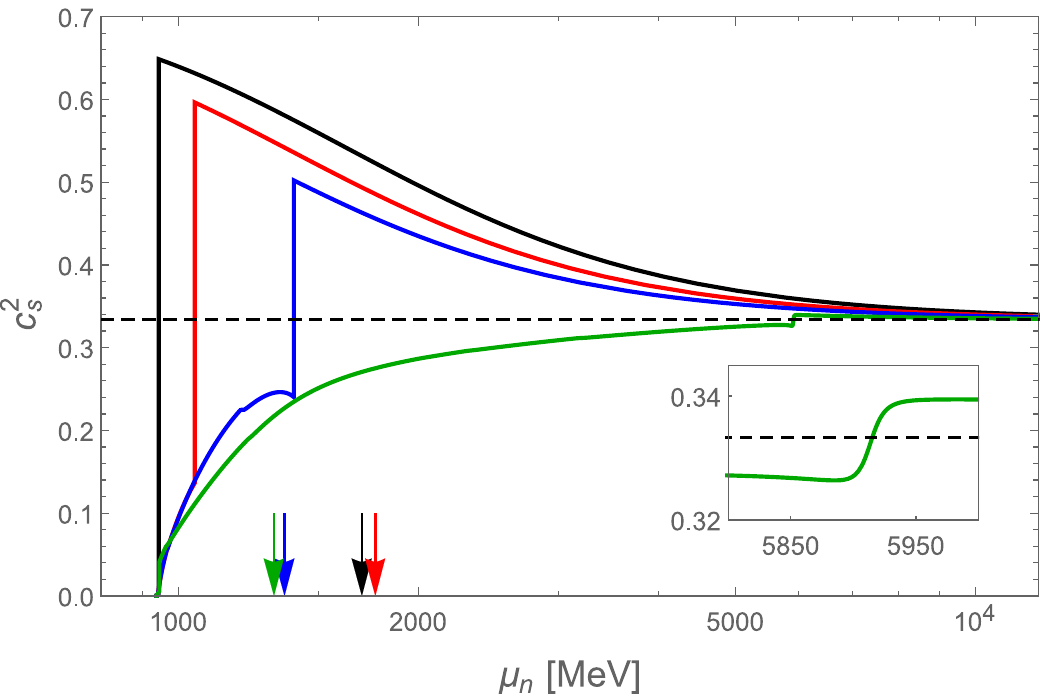}
\caption{Speed of sound squared as a function of the neutron chemical potential for the four cases of Fig.\ \ref{fig:MNmun}, showing only the stable branches. Colors correspond to the colors of Fig.\ \ref{fig:MNmun}, i.e. the Dirac mass at saturation increases from black to red to blue to green. The large discontinuities in the black, red, and blue curves indicate the chiral phase transition, while the green curve has a chiral crossover, as the zoom-in proves. The arrows mark the chemical potentials in the center of the most massive star of each case, and the horizontal dashed line marks the conformal value, $c_s^2=1/3$, that is attained asymptotically by all curves. }
\label{fig:cs}
\end{center}
\end{figure}

We show the speed of sound squared $c_s^2$ for the four parameter sets of the previous subsection in Fig.\ \ref{fig:cs}. This figure contains various interesting aspects. First, we see that all curves approach the conformal limit $c_s^2=1/3$, as already suggested by the analytical calculation in Sec.\ \ref{sec:sound}. While that calculation was performed for symmetric nuclear matter without strangeness, here we see that the conformal limit is also assumed asymptotically in the electrically neutral, beta-equilibrated case including strange matter. As pointed out in Sec.\ \ref{sec:sound}, the nonzero value of the vector meson self-coupling $d$ is crucial for this behavior. Second, the zoom-in shows that the lower right panels of Figs.\ \ref{fig:MNmun} and \ref{fig:fractions} indeed contain a smooth chiral crossover: the speed of sound -- containing a second derivative of the free energy -- is continuous and smooth.  

Third, and perhaps most importantly, let us comment on the behavior of the speed of sound in the intermediate density regime, relevant for neutron star matter. It is striking that in
the cases of a first-order chiral transition the speed of sound {\it increases} through the discontinuity as we move towards large densities. Even in the case of the crossover this tendency is retained; through the sharp crossover the speed of sound is increased from just below to just above the conformal limit. (We have checked that there are parameter sets where $c_s^2>1/3$ before the sharp crossover, i.e., this is not a generic feature.) 
The large speed of sound in our chirally restored phase is somewhat surprising if we have in mind perturbative QCD, which predicts $c_s^2<1/3$ where it is applicable. We should emphasize that our model is not asymptotically free. Even though the conformal limit is approached asymptotically, interactions still play a role in this limit. Therefore, we cannot expect to reproduce this prediction of perturbative QCD. At intermediate densities, QCD is strongly coupled and we have no first-principle results for the speed of sound of quark matter. Therefore, our result is not in any contradiction with QCD. Another reason to expect a smaller speed of sound in the chirally restored phase might be the increase in degrees of freedom as we cross the phase transition. While this tends to soften the equation of state, i.e., to decrease the speed of sound, there are at least two opposing effects that, in our model, turn out to dominate the behavior. Namely, the near-masslessness of the degrees of freedom in the chirally restored phase should indeed contribute to an increase of the speed of sound, and, of course, the form of the interactions plays an important role, which is not easy to disentangle from the other effects. A speed of sound of quark matter above the conformal limit has also been observed in resummed perturbation theory \cite{Fujimoto:2020tjc} and in the color-flavor locked phase  \cite{Traversi:2021fad}. In fact, it has been shown that no exotic degrees of freedom are necessary in order to generate a speed of sound that surpasses its asymptotic conformal limit. Rather, a peak in the speed of sound of homogeneous matter naturally emerges in the transition from a phase with broken chiral symmetry to one with a gapped Fermi surface \cite{Hippert:2021gfs}.

\begin{figure}[t]
\begin{center}
\includegraphics[width=0.49\textwidth]{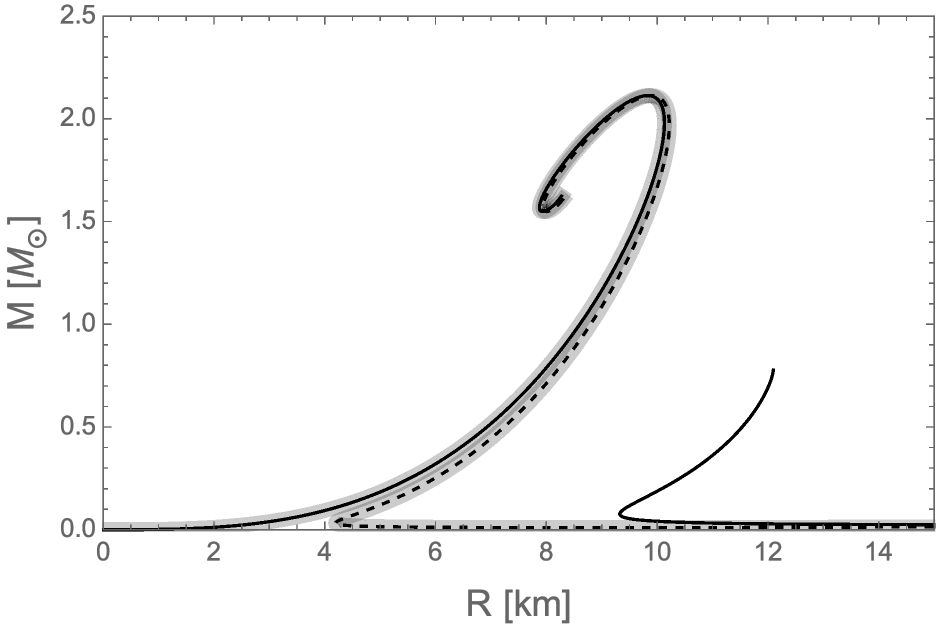}
\includegraphics[width=0.49\textwidth]{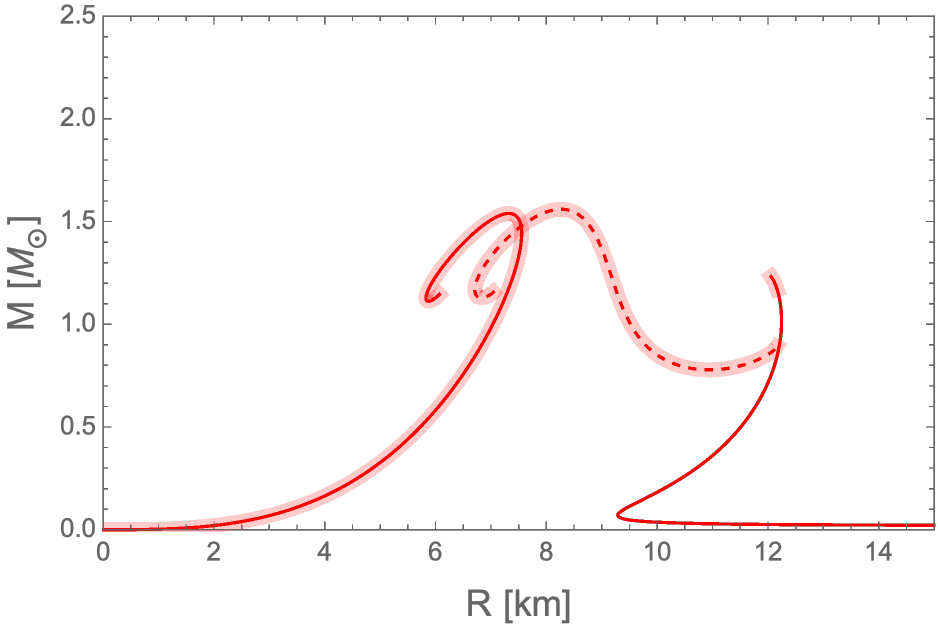}

\includegraphics[width=0.49\textwidth]{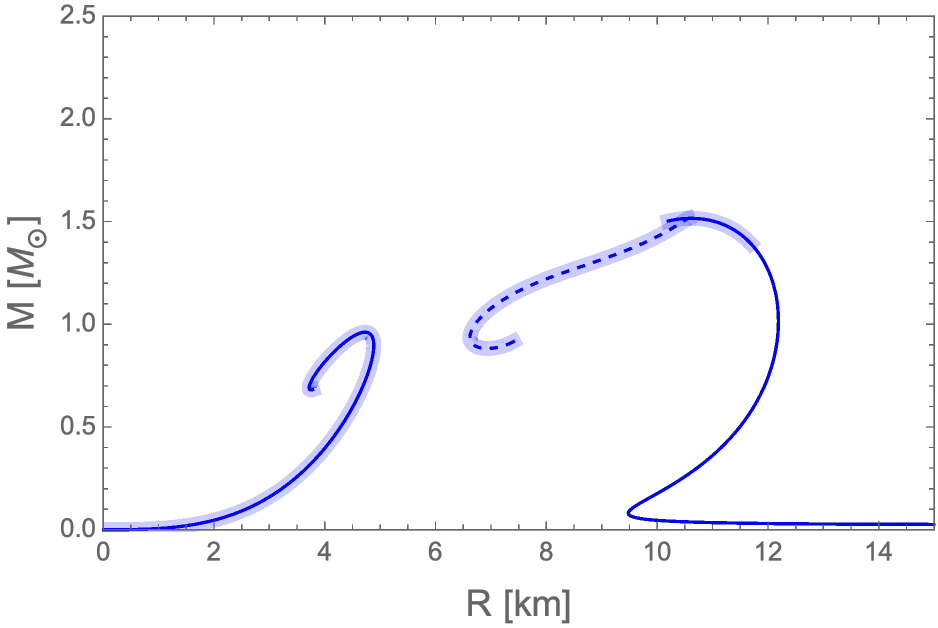}
\includegraphics[width=0.49\textwidth]{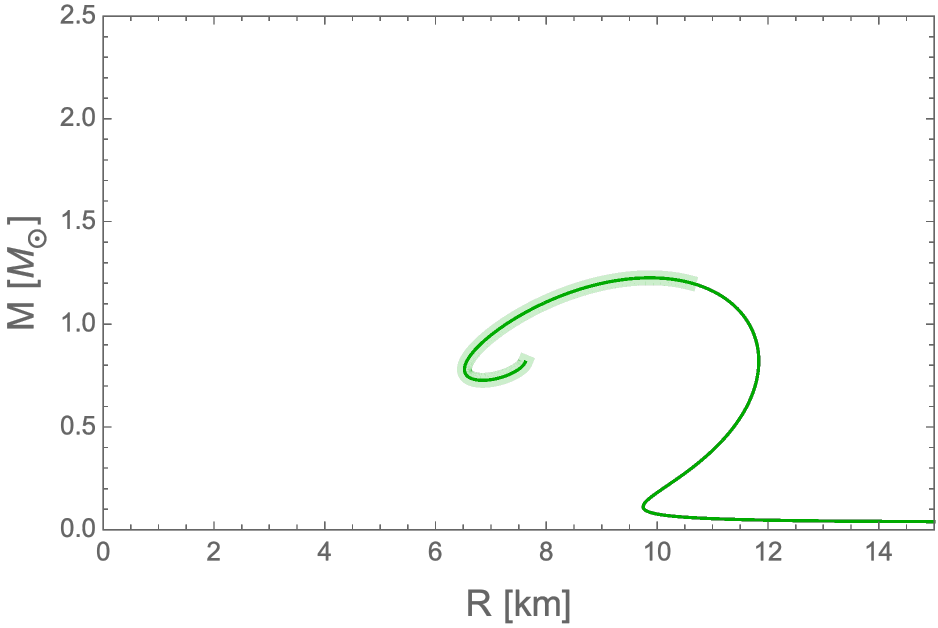}
\caption{Mass-radius curves of quark stars (curves reaching back to the origin), hybrid stars (dashed), and neutron stars for the four cases of Fig.\ \ref{fig:MNmun}.  The shaded bands mark the stars containing strangeness. Only the upper left panel is in accordance with the heaviest known neutron star. (Radius constraints must be ignored here since we have not included a crust, which would change radii, but not the maximal masses, significantly.) The lower right panel corresponds to a parameter set with a chiral crossover and thus only has a single class of stars. }
\label{fig:MR}
\end{center}
\end{figure}

The speed of sound is a measure for the stiffness of matter, and we expect stiff matter to give rise to large neutron star masses. This connection is borne out in the mass-radius curves shown in Fig.\ \ref{fig:MR}. They are computed by inserting the equation of state $P(\epsilon)$, with pressure $P=-\Omega$ and energy density $\epsilon$ from Eqs.\ (\ref{Omega}) and (\ref{epsilon}), into the so-called Tolman-Oppenheimer-Volkoff equations \cite{PhysRev.55.364,tolman,PhysRev.55.374}, which describe a static, spherically symmetric matter configuration in general relativity. By choosing the central pressure as a boundary condition and solving the differential equations numerically one obtains the mass and radius of the star. Varying the central pressure generates a mass-radius curve, representing all possible stars for a given equation of state. 

In Fig.\ \ref{fig:MR} we show three different classes of stars, which are best explained with the help of the free energy in Fig.\ \ref{fig:Om}: ``neutron stars", i.e., stars made entirely of baryonic matter, 
probe the chirally broken branch of our solution. Their maximal central pressure is given by the phase transition point ($\mu_n\simeq 1.4\, {\rm GeV}$ in Fig.\ \ref{fig:Om}) if only stable baryonic matter is considered. In the mass-radius plots we have traced the neutron star branch beyond the transition point into the spinodal region, following the (now metastable) chirally broken solution. Importantly, this spinodal region ends at some point, which corresponds to the end points of the neutron star curves in Fig.\ \ref{fig:MR}. In an approach using different models for quark and hadronic matter the metastable branch would continue to arbitrarily large densities and no prediction for the endpoint in the mass-radius curve can be made. This metastable neutron star segment can be of astrophysical relevance since it is made of two-flavor nuclear matter (entirely in the upper left panel and for a large part in the upper right panel). Therefore, it is conceivable that it survives for non-microscopic times since the conversion to strange quark matter would require the injection of strangelets.

If we follow the thermodynamically stable branches through the phase transition, we  branch off of the neutron star curve by following the chirally restored branch. We obtain hybrid stars, shown by the dashed curves in Fig.\ \ref{fig:MR}, i.e., stars with a chirally broken mantle and a chirally restored core. This gives rise to the possibility of ``twin stars", stable stars with the same mass but different radii \cite{Benic:2014jia}. Twins both having  thermodynamically stable matter -- one neutron star, one hybrid star -- are (barely) realized in the upper right panel. However, our results also suggest the existence of twins where one star is made of metastable hadronic matter and its hybrid twin containing a strange quark matter core (upper panels). In all mass-radius plots we have included segments that are expected to be unstable with respect to radial oscillations of the star \cite{1966ApJ...145..505B,glendenningbook}. Therefore, for instance, the lower left panel does not allow for twin stars because the entire hybrid branch is expected to be unstable. 

We also show the mass-radius curves of ``quark stars" made entirely out of chirally restored matter in our model. To this end, we follow the chirally restored solution in Fig.\ \ref{fig:Om} backwards until the pressure (and thus the free energy density) is zero. In the three cases considered here where this construction is possible, this includes a metastable segment of the solution, towards low densities, similar to the metastable neutron stars just discussed, where the metastable matter sits at high densities. 
There are parameter regions where the metastable segment does not reach back to zero pressure, which results in quark matter only appearing in hybrid stars, and not also in a separate branch of quark stars. 
There are also parameter regions where the chirally restored branch is stable all the way down to zero pressure, which we can interpret as a realization of the strange quark matter hypothesis \cite{PhysRevD.30.272,PhysRevD.4.1601}. We shall come back to this possibility -- and identify the region in the parameter space where it is realized -- in the subsequent section. 

In the calculation of the mass-radius curves we have not included any mixed phase at the chiral phase transition. A mixed phase layer in the star would smoothen the cusp-like transition from the neutron star branch to the hybrid star branch, but otherwise is not expected to change the results significantly. Moreover, we have not included a crust but rather used the homogeneous phases of our model down to the lowest densities. This simplification has a large effect on the radii of the stars. A crust would generate a much larger layer of matter with an average density below saturation density and can be expected to correct the radii to much larger values (see for instance Ref.\ \cite{Kovensky:2021kzl}), with the exception of the quark stars, where only a small crust is expected (see for instance Ref.\ \cite{Zapata:2021hwt}). Importantly, however, the inclusion of a crust and its precise properties are not expected to change the maximal mass of the given dense matter equation of state \cite{Kovensky:2021kzl}. Therefore, the radii in Fig.\ \ref{fig:MR} should not be taken too seriously, and we should thus not attempt to compare these results to the latest data for neutron star radii, and neither to constraints for the tidal deformability, which is strongly influenced by the radius of the star. However, the maximal mass of our mass-radius curves {\it can} be taken seriously. As a consequence, we see that only the upper left panel corresponds to an equation of state allowed by the existence of a 2.1-solar mass star \cite{NANOGrav:2019jur,Fonseca:2021wxt}. In particular, the scenario with the chiral crossover (lower right panel) gives rise to very low masses and thus is in contradiction with astrophysical data. These observations reflect the behavior of the speed of sound in Fig.\ \ref{fig:cs}: heavy stars are possible for large speeds of sound, and the largest mass is obtained for the case with the earliest chiral phase transition such that the stiff chirally restored phase constitutes a large volume fraction of the heaviest stars. This is in line with recent discussions suggesting the necessity for a non-monotonic behavior of the speed of sound in order to meet astrophysical constraints \cite{Tews:2018kmu,Annala:2019puf,Altiparmak:2022bke}. While in many approaches, either purely baryonic or in connection with a separate quark matter model, the maximum of the speed of sound is reached in the baryonic phase it has also been argued that this behavior may be generated by the  so-called quarkyonic phase \cite{McLerran:2018hbz}. In contrast, our results suggest that the peak of the speed of sound may well appear in the quark matter phase, while the baryonic phase exhibits sound speeds below the conformal limit.

\subsection{Parameter-independent conclusions}
\label{sec:paraspace}

\begin{figure} 
\begin{center}
\includegraphics[width=0.49\textwidth]{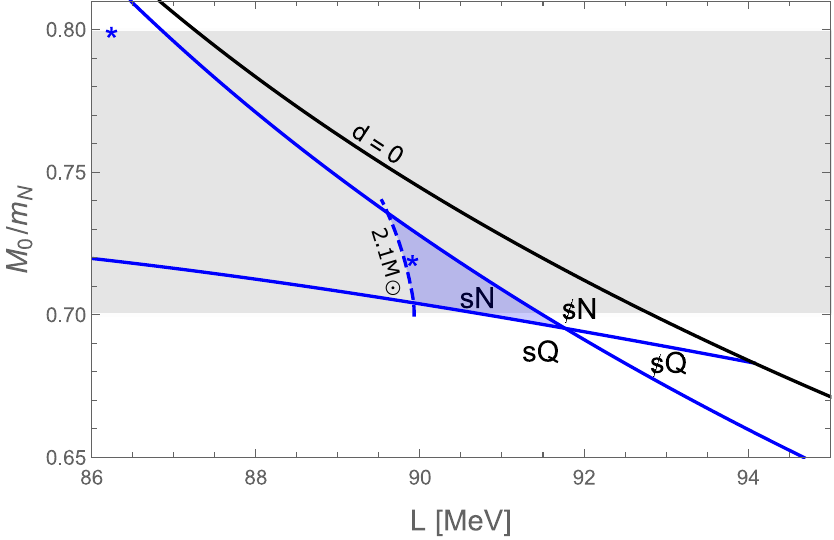}
\includegraphics[width=0.49\textwidth]{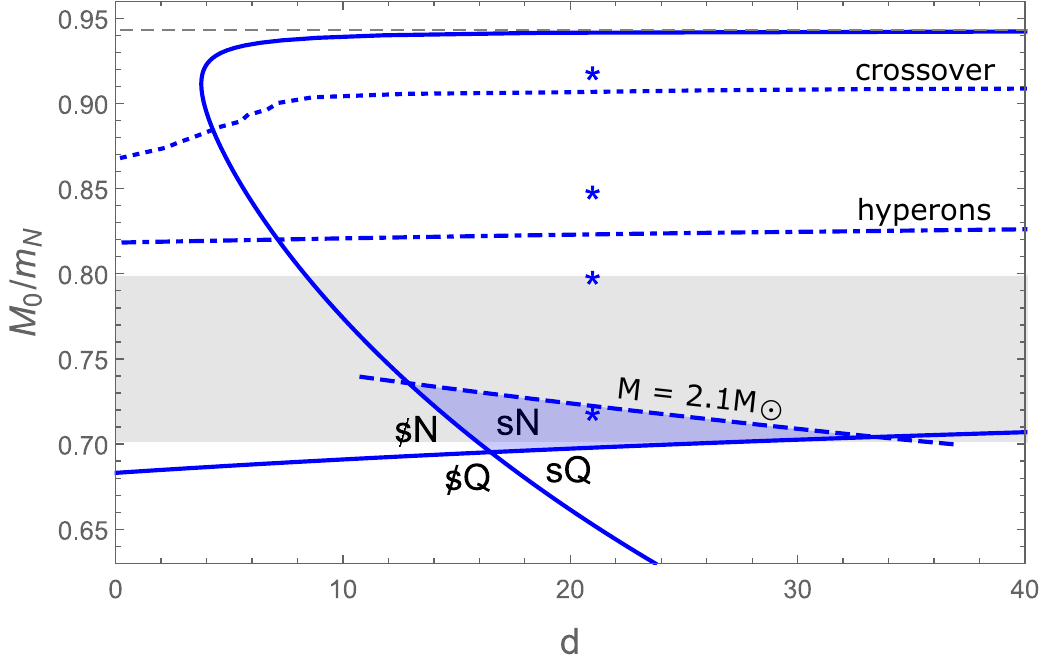}

\includegraphics[width=0.49\textwidth]{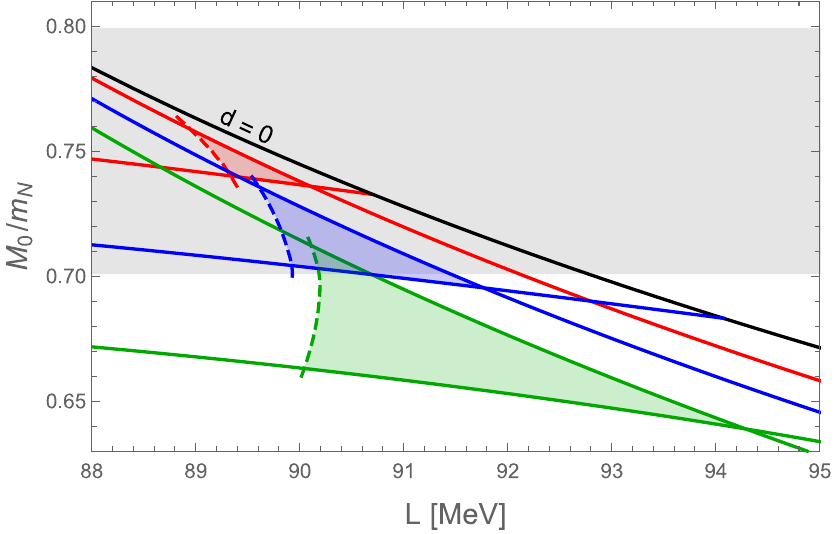}
\includegraphics[width=0.49\textwidth]{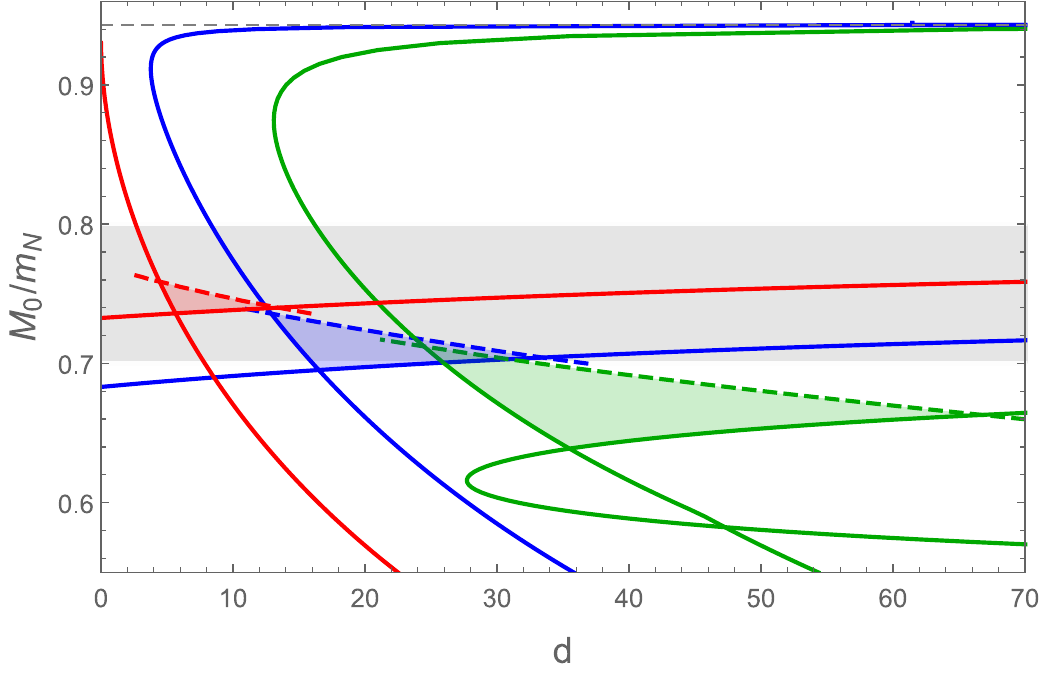}
\caption{{\it Upper panels:} Distinct regions in the $M_0$-$L$ and $M_0$-$d$ planes for $K=250\, {\rm MeV}$ and ${\cal U} = -50\, {\rm MeV}$. ``s" (``$\slashed{\rm s}$") labels regions with (without) strangeness at asymptotic densities, ``N" (``Q") labels regions where nuclear matter (quark matter) is preferred at zero pressure. In the shaded triangular region maximal masses of (hybrid) stars of more than 2.1 solar masses are reached, in addition to having asymptotic strangeness and nuclear matter being stable at zero pressure. In the right panel, the dashed-dotted (almost horizontal) curve divides the region where hyperons appear before the chiral transition (towards large $M_0$) from the region where strangeness only appears in the chirally restored phase (towards small $M_0$). Above the dotted line the chiral transition is a crossover. The grey shaded band
in both panels is the empirically preferred  regime for $M_0$, and the thin horizontal dashed line in the right panel marks the upper limit of $M_0$ according to Fig.\ \ref{fig:M0Lspace}. The asterisks  correspond to the parameter choices in Figs.\ \ref{fig:MNmun} -- \ref{fig:MR} (in the left panel only two of them lie in the shown range). {\it Lower panels:} Blue lines as in the upper panels, now with added curves for ${\cal U}=-30\, {\rm MeV}$ (green) and ${\cal U}=-70\, {\rm MeV}$ (red). }
\label{fig:M0Ld}
\end{center}
\end{figure}

We have seen that our model allows for qualitatively different scenarios regarding the chiral phase transition, with different thermodynamic properties and different properties of compact stars. We now intend to determine the region in parameter space  where our model is useful and realistic. For simplicity we keep the incompressibility at saturation fixed to $K=250\, {\rm MeV}$, and vary the Dirac mass at saturation $M_0$, the slope parameter $L$, and the hyperon potential ${\cal U}$. We present our results in the 
$M_0$-$L$ plane, making the connection to Fig.\ \ref{fig:M0Lspace}. It is useful to consider also the $M_0$-$d$ plane for an alternative representation. For a given pair $(M_0,d)$ one can always compute the more physical pair $(M_0,L)$. 

Our results are shown in Fig.\ \ref{fig:M0Ld}. 
Let us first focus on the upper panels, which are obtained with the choice ${\cal U}=-50\, {\rm MeV}$, to explain and interpret the various curves. 
\begin{itemize}
\item {\it Asymptotic strangeness.} For our main goal to describe strange quark matter with our chirally restored phase we need to check in which cases there is strangeness at asymptotically large densities. (As we have seen in Fig.\ \ref{fig:fractions}, if strangeness survives asymptotically, it tends to be present right after the phase transition as well.) The line in the parameter space that separates the region with asymptotic strangeness from the one without can be calculated with the help of an expansion similar to the asymptotic expansion employed in Appendix \ref{app:symmetry}. The ansatz for the solution of the stationarity equations used in this appendix led to conditions for the coupling constants, guaranteeing {\it flavor-symmetric} asymptotic strangeness. The weaker condition of the existence of asymptotic strangeness is found by the ansatz $\rho\simeq \rho_\infty \mu_n$, $\mu_e\simeq \mu_{e,\infty}\mu_n$ and all other condensates as in Eq.\ (\ref{ansatz}). This ansatz leads to a set of stationarity equations for the coefficients of the leading-order terms $\omega_\infty,\rho_\infty,\phi_\infty,\mu_{e,\infty}$, which can easily be solved numerically. Then, for instance at a fixed $d$, we can determine the value of $M_0$ at which a strange degree of freedom first sets in asymptotically, and repeating the procedure for many values of $d$ gives a curve in the $M_0$-$d$ plane and thus also in the $M_0$-$L$ plane, shown as a blue solid curve, where regions with and without asymptotic strangeness are labeled by  ``s" and ``$\slashed{\rm s}$". 

\item {\it Stability of nuclear matter at zero pressure.} If our chirally restored phase is favored at zero pressure, it prevails for all nonzero densities and the main purpose of the model, to develop a unified approach in the vicinity of the quark-hadron transition, is not realized. Therefore we need to identify the parameter region in which nuclear matter is the favored phase at zero pressure. We can compute the line that bounds this region by computing the points in the $M_0$-$d$ plane at which  chirally restored, zero-pressure matter sits exactly at $\mu_n=m_N$, where the second-order onset of charge neutral, beta-equilibrated nuclear matter occurs. If it sits at larger $\mu_n$, as in Fig. \ref{fig:Om} and all parameter sets of Sec.\ \ref{sec:4sets}, there is a chiral transition and we denote this case in Fig.\ \ref{fig:M0Ld} by ``N"; if it sits at lower $\mu_n$ there is no chiral transition and we denote this case by ``Q" since it suggests that quark matter is absolutely stable. Together 
with the criterion for asymptotic strangeness we find four regions: sN, sQ, $\slashed{\rm s}$N, $\slashed{\rm s}$Q. (The $M_0$-$L$ plane additionally has the region of negative $d$, which we do not consider.) {\it For our purpose, the sN region -- asymptotic strangeness and absolutely stable nuclear matter -- is the most relevant.}

\item {\it Realistic neutron stars.} On the blue dashed curve the maximal mass of a hybrid star is exactly $2.1\, M_\odot$, heavier stars are sitting to the right (upper left panel) or below (upper right panel) this curve. We have restricted this curve to the sN region and only indicated that it also extends into the the sQ region (where there are no hybrid stars, i.e. the maximal mass is reached by a quark star) and into the $\slashed{\rm s}$N region. The resulting window in the sN region containing stars with maximal masses compatible with astrophysical data is shaded in blue. One of the four parameter sets of Sec.\ \ref{sec:4sets}, indicated by asterisks, lies in that region. {\it We see that the shaded region is compatible with the empirical constraints for $M_0$, and that it defines a remarkably narrow range in $L$.} As a measure for the largest possible mass of the star inside the triangular region we have also computed the mass at the tip of the triangle opposite of the dashed curve and found  $M\simeq 2.28\, M_\odot$, i.e. if a star with a larger mass than that value was measured, our shaded region would disappear. [For the two additional parameter sets in the lower panels, these values are $M\simeq 2.36\, M_\odot$ (green) and $M\simeq  2.23 \, M_\odot$ (red).]

\item {\it Appearance of hyperons.} Parameter choices above the dashed-dotted curve in the upper right panel lead to the appearance of hyperons. More precisely, to plot this curve we have for each $d$  determined the $M_0$ at which we first see the appearance of (any) strange degrees of freedom just below the chiral phase transition, i.e., at the lower density of the density jump. We find that hyperons only appear for very large values of $M_0$. Although the boundaries of the grey band $M_0\simeq (0.7-0.8)m_N$ should not be taken as sharp constraints, it is unlikely that $M_0$ assumes such a large value. Perhaps more importantly, hyperons only appear in a region where the maximal masses of compact stars are well below two solar masses. This observation puts our results into the context of the ``hyperon puzzle" \cite{Tolos:2020aln}: while hyperons are expected to appear at sufficiently large chemical potentials they tend to soften the equation of state and thus render large masses of neutron stars impossible. This is exactly what our model shows, and, importantly, within the same model a solution is suggested, namely {\it the appearance of a stiff chirally restored phase before a potential hyperon onset, allowing for  sufficiently heavy hybrid stars.} 

\item {\it Crossover.} The dotted line at even larger $M_0$ marks the change from a first-order chiral transition to a crossover. In other words, below that line there is a multivalued solution of the stationarity equations at high densities, and for each $d$ we have determined the $M_0$ where the solution turns into a single-valued curve. As already suggested by Fig.\ \ref{fig:MR}, the scenario of a chiral crossover is -- within our model -- incompatible with realistic maximal masses of compact stars. 

\end{itemize}

In the lower two panels of Fig.\ \ref{fig:M0Ld} we have added the curves for two different values of the hyperon potential, ${\cal U} = -30\, {\rm MeV}$ and ${\cal U} = -70\, {\rm MeV}$. To avoid too much cluttering we do not show the hyperon onset and crossover lines for these cases, but we have checked that they are also above the grey band, i.e., in an empirically unfavored region.  In the lower right panel we see that the line separating absolutely stable nuclear matter from the region where the strange quark matter hypothesis is realized (i.e., ``N" from ``Q") looks qualitatively different for larger (less negative) hyperon potentials. This gives rise to a second, disconnected sN region, which, however, is disfavored due its incompatibility with the empirical constraints for $M_0$.  We also observe  that for less negative values the shaded area leaves the grey band. 
Sufficiently heavy stars still exist in the grey band, but not in conjunction with asymptotic strangeness, which tends to disappear if ${\cal U}$ is made less negative or even positive. As we mentioned at the end of Sec.\ \ref{sec:couplings}, in the realistic parameter regime the hyperon potentials are effectively only relevant for the chirally restored phase, fixing the interactions between light degrees of freedom because actual hyperons do not appear in this parameter regime. If, on the other hand, we go to even more negative ${\cal U}$, the triangular region itself becomes smaller and smaller as it moves to larger values of $M_0$ and smaller $L$. As a consequence, the most important conclusion from the lower plots is that the prediction for the value of $L$ is not altered much by allowing the hyperon potential to vary. The lower left panel suggests that {\it independently of the value of the hyperon potential the allowed region of $L$ turns out to be $L\simeq (88-92)\, {\rm MeV}$.} This is a remarkably narrow range, which can be expected to become somewhat larger by exhausting the remaining uncertainties in the incompressibility $K$ and the symmetry energy $S$.

\section{Summary and outlook}
\label{sec:summary}

We have discussed cold and dense matter undergoing a chiral phase transition within a nucleon-meson model. The main idea has been to include strange baryonic degrees of freedom in the Lagrangian, not necessarily to account for hyperons, which may or may not be favored, but to create a chirally restored phase that resembles strange quark matter. We have pointed out that it is possible to choose the parameters of the model such that flavor-symmetric matter is obtained at ultra-high densities, as expected from asymptotically dense three-flavor quark matter in QCD. However, in this parameter regime the model does not produce compact stars with masses that meet the astrophysical constraints. Therefore, we have  
mainly explored a parameter region which is not flavor-symmetric asymptotically, but still has nonzero strangeness for large densities and a speed of sound that approaches the conformal limit, as expected from QCD. 

Within this parameter region, we have shown that  qualitatively different scenarios are possible regarding the chiral phase transition (first order vs.\ crossover) and the onset of strangeness (within the baryonic phase as hyperons vs.\ only in the chirally restored phase). Requiring the model to produce compact stars of at least 2.1 solar masses and the correct saturation properties of symmetric nuclear matter disfavors a chiral crossover and the appearance of hyperons. The heaviest stars in the model turn out to be hybrid stars, which can be traced back to a large speed of sound in the chirally restored phase, which peaks just after the chiral phase transition. Furthermore, putting together low-density and astrophysical constraints we have shown that the poorly known slope parameter of the symmetry energy is narrowed down to about $L\simeq (88-92)\, {\rm MeV}$. Due to the phenomenological nature of the model and the simplifications we have made, these numbers should of course be taken with some care.

The main motivation for developing this setup was to provide a unified approach for both quark and hadron phases which enables us to consistently compute properties of matter in the vicinity of the chiral phase transition, such as the surface tension, the free energy of a mixed phase, or the possible existence of an inhomogeneous chiral condensate, for instance in the form of a chiral density wave. Especially in view of the significance of (global) electric charge neutrality in a neutron star, the inclusion of strangeness has been a step forward because starting with non-strange baryonic degrees of freedom leaves us with no negative charge carriers (except for leptons) in the chirally symmetric phase. These applications of the model are thus natural directions for the future. 

It would also be interesting to further improve the model itself. For instance, we have neglected the hidden-strangeness scalar condensate, such that all baryon masses are generated by the non-strange chiral condensate. Also, one may include small explicit baryon masses, which we set to zero, such that the chiral symmetry was explicitly broken only through the potential of the scalar field and a non-$SU(3)$ symmetric choice of the baryon-meson coupling constants. We have also restricted ourselves to zero temperatures, and extensions to finite temperatures, desirably going beyond the mean-field approximation, would be interesting and relevant for applications to the mergers of compact stars in the presence of a quark-hadron transition \cite{Most:2019onn,Blacker:2020nlq}. 

Moreover, applications of 
our idea to related models are possible, for instance by including strange baryonic degrees of freedom (and their chiral partners) into the extended linear sigma model of Refs.\ \cite{Detar:1988kn,Zschiesche:2006zj,Heinz:2013hza}. It is also conceivable to include Cooper pairing, both in the chirally broken and the chirally restored phases, and it would be interesting to see whether a version of the color-flavor locked phase at high densities, where all fermionic degrees of freedom participate in pairing \cite{Alford:1998mk,Alford:2007xm}, can be constructed. It would then be possible to compute for instance the surface tension in the presence of Cooper pairing consistently within a single model. Or, considering the case of a crossover, the model might be able to provide a realization of the quark-hadron continuity in the sense suggested by Ref.\ \cite{Schafer:1998ef}. Finally, our setup can be used for studying a possible quarkyonic phase, which has been predicted to occur in QCD at a large number of colors $N_c$ and may survive for $N_c=3$ \cite{McLerran:2007qj}. This phase was for instance constructed in a model that includes both quark and hadronic degrees of freedom \cite{Cao:2020byn} (besides other approaches \cite{McLerran:2018hbz,Kovensky:2020xif,Margueron:2021dtx}). It would be interesting to see whether our more unified approach might be able to show a transition from baryonic through  quarkyonic  to  quark matter.

\begin{acknowledgments}
We thank A.\ Haber and D.\ Rischke for helpful comments and discussions. 
E. S. F. is partially supported by CAPES (Finance Code 001), CNPq, FAPERJ, and INCT-FNA (Process No. 464898/2014-5). 
R.M. thanks CNPq for financial support.
\end{acknowledgments}

\appendix

\section{Chiral setup}
\label{app:su3}

In this appendix we review the foundations of our model within the framework of an $SU(3)\times SU(3)$ chiral approach. This discussion makes explicit which mesonic degrees of freedom we have omitted and which assumptions we have made for the structure of the interaction terms, which is useful to keep in mind for potential extensions in the future. It also provides relations between the baryon-meson coupling constants, some of which we employ in the main part, besides guidance from phenomenology to fix them. Our baryonic degrees of freedom are usually parametrized in the baryon octet as
\be
B =\left(\begin{array}{ccc} \displaystyle{\frac{\Sigma^0}{\sqrt{2}} + \frac{\Lambda}{\sqrt{6}}} & \Sigma^+ & p \\ \Sigma^- &
\displaystyle{-\frac{\Sigma^0}{\sqrt{2}} + \frac{\Lambda}{\sqrt{6}}}  & n  \\ \Xi^- & \Xi^0 & \displaystyle{-\sqrt{\frac{2}{3}}\Lambda} 
\end{array}\right) \, , 
\ee
and the kinetic part of the baryonic Lagrangian can be written as $\Tr[\bar{B}i\gamma^\mu\partial_\mu B]$. 
The scalar and pseudoscalar meson nonets are summarized in the field $\Phi  =  S+iP = T_a(\sigma_a+i\pi_a)$, where $T_a=\lambda_a/2$ for $a=0,\ldots, 8$, with  the Gell-Mann matrices $\lambda_a$ for $a=1,\ldots,8$ and $\lambda_0 = \sqrt{2/3}\,{\bf 1}$. This is usually reparametrized as 
\begin{subequations}
\bea
S=T_a\sigma_a &=& \frac{1}{\sqrt{2}}\left(\begin{array}{ccc} \displaystyle{\frac{a_0^0}{\sqrt{2}} + \frac{\sigma_8}{\sqrt{6}}+\frac{\sigma_0}{\sqrt{3}}} & a_0^+ & \kappa^+ \\ a_0^- &
\displaystyle{-\frac{a_0^0}{\sqrt{2}} + \frac{\sigma_8}{\sqrt{6}}+\frac{\sigma_0}{\sqrt{3}}}  & \kappa^0  \\ \kappa^- & \bar{\kappa}^0 & \displaystyle{-\sqrt{\frac{2}{3}}\sigma_8+\frac{\sigma_0}{\sqrt{3}}} 
\end{array}\right) \, , \\[2ex]
P=T_a\pi_a &=& \frac{1}{\sqrt{2}}\left(\begin{array}{ccc} \displaystyle{\frac{\pi^0}{\sqrt{2}} + \frac{\pi_8}{\sqrt{6}}+\frac{\pi_0}{\sqrt{3}}} & \pi^+ & K^+ \\ \pi^- &
\displaystyle{-\frac{\pi^0}{\sqrt{2}} + \frac{\pi_8}{\sqrt{6}}+\frac{\pi_0}{\sqrt{3}}}  & K^0  \\ K^- & \bar{K}^0 & \displaystyle{-\sqrt{\frac{2}{3}}\pi_8+\frac{\pi_0}{\sqrt{3}}} 
\end{array}\right)   \, .
\eea
\end{subequations} 
One may now construct the potential up to a given order in $\Phi$ systematically. For instance, up to fourth order \cite{Lenaghan:2000ey}, 
\be \label{UPhi}
U(\Phi) = m^2\Tr[\Phi^\dag\Phi]+\lambda_1(\Tr[\Phi^\dag\Phi])^2+\lambda_2\Tr[(\Phi^\dag\Phi)^2]-c({\rm det}\,\Phi^\dag+{\rm det}\,\Phi)-\Tr[H(\Phi^\dag+\Phi)] \, ,
\ee
with parameters $m^2, \lambda_1, \lambda_2$ for the quadratic and quartic contributions, $c$ for the chiral anomaly term and a matrix $H$ for a small explicit symmetry breaking. 
In the scalar sector, one can trade $\sigma_0$ and $\sigma_8$ for 
non-strange and strange scalar fields by the transformation 
\be \label{trafo}
\left(\begin{array}{c} \sigma \\ \zeta\end{array}\right) = \frac{1}{\sqrt{3}}\left(\begin{array}{cc} \sqrt{2} & 1 \\ 1 & -\sqrt{2} \end{array}\right)\left(\begin{array}{c} \sigma_0 \\ \sigma_8\end{array}\right) \, . 
\ee
Omitting all other scalar fields results in  
$S = \frac{1}{2}{\rm diag}(\sigma,\sigma,\sqrt{2}\zeta)$. As explained in the main text we further simplify this by omitting the scalar field $\zeta$. The pseudoscalar nonet $P$ is not directly relevant because we assume none of these fields to condense, and our mean-field approach 
ignores the fluctuations. It is only indirectly used by fitting one of the parameters of the meson potential (\ref{U}) to the pion mass. 
Our potential thus effectively only depends on $\sigma$, which is a drastic simplification of the full potential (\ref{UPhi}). However, we have included terms of higher order than 4 in $\sigma$, to make the connection with the previous (non-strange) version of our baryon-meson model \cite{Fraga:2018cvr,Schmitt:2020tac}.  

The vector meson nonet can be parametrized as 
\bea
V_\mu=T_a\omega^a_\mu &=& \frac{1}{\sqrt{2}}\left(\begin{array}{ccc} \displaystyle{\frac{\rho^0_\mu}{\sqrt{2}} + \frac{\omega_\mu}{\sqrt{2}}} & \rho^+_\mu & K^{*+}_\mu \\ \rho^-_\mu &
\displaystyle{-\frac{\rho^0_\mu}{\sqrt{2}} + \frac{\omega_\mu}{\sqrt{2}}}  & K^{*0}_\mu  \\[3ex] K^{*-}_\mu & \bar{K}^{*0}_\mu & \phi_\mu 
\end{array}\right) \, , 
\eea
where $\omega_\mu$ and $\phi_\mu$ are defined by the same transformation as used in Eq.\ (\ref{trafo}) for the scalar mesons,
\be
\left(\begin{array}{c} \omega_\mu \\[1ex] \phi_\mu \end{array}\right) = \frac{1}{\sqrt{3}}\left(\begin{array}{cc} \sqrt{2} & 1 \\[1ex] 1 & -\sqrt{2} \end{array}\right)\left(\begin{array}{c} \omega^0_\mu \\[1ex] \omega^8_\mu\end{array}\right) \, .
\ee
Keeping only the fields $\omega_\mu$, $\phi_\mu$, $\rho^0_\mu$, the matrix $V_\mu$ becomes diagonal, and we can write down the two quartic structures 
\be \label{d1d2}
d_1(\Tr[V_\mu V^\mu])^2 + d_2\Tr[(V_\mu V^\mu)^2] = \frac{d_1}{4}(\omega_\mu\omega^\mu+\rho^0_\mu\rho_0^\mu+\phi_\mu\phi^\mu)^2 + \frac{d_2}{8}\left[(\omega_\mu\omega^\mu)^2+(\rho^0_\mu\rho_0^\mu)^2+6\omega_\mu\omega^\mu\rho_\nu^0\rho^\nu_0\right] \, .
\ee
In the main text we work for simplicity with $d_2=0$ (and denote $d=d_1$). 
For a more complete study of vector meson self-interactions in a chiral approach, including axial-vector mesons and derivative interactions with three meson fields, see for instance Refs.\ \cite{Parganlija:2010fz,Parganlija:2012fy}.

Next, let us discuss the baryon-meson interactions. For the scalar sector, and temporarily including the $\zeta$ field, the chirally invariant structures are 
\bea
&&A_1\Tr[\bar{B}SB]+A_2\Tr[\bar{B}BS]+A_3\Tr[\bar{B}B]\Tr[S] \non[2ex]
&&=
g_{N\sigma}(\bar{n}\sigma n+\bar{p}\sigma p) 
+ g_{N\zeta}(\bar{n}\zeta n+\bar{p}\zeta p)+g_{\Sigma\sigma}(\bar{\Sigma}^0\sigma\Sigma^0+\bar{\Sigma}^+\sigma\Sigma^++\bar{\Sigma}^-\sigma\Sigma^-)
+g_{\Sigma\zeta}(\bar{\Sigma}^0\zeta\Sigma^0+\bar{\Sigma}^+\zeta\Sigma^++\bar{\Sigma}^-\zeta\Sigma^-) \non[2ex]
&&+g_{\Lambda\sigma}\bar{\Lambda}\sigma\Lambda
+g_{\Lambda\zeta}\bar{\Lambda}\zeta\Lambda+g_{\Xi\sigma}(\bar{\Xi}^0\sigma\Xi^0+\bar{\Xi}^-\sigma\Xi^-)
+g_{\Xi\zeta}(\bar{\Xi}^0\zeta\Xi^0+\bar{\Xi}^-\zeta\Xi^-) \, .
\eea
We have introduced 8 coupling constants, which all are linear combinations the 3 independent parameters $A_1$, $A_2$, $A_3$. Therefore, one can choose 3 independent couplings, and the chiral structure fixes the other 5. In our approximation, where we omit the $\zeta$, we have 4 coupling constants and thus, if we wanted to respect the structure given by chiral symmetry, we can choose three of them freely, say $g_{N\sigma}$, $g_{\Sigma\sigma}$, $g_{\Lambda\sigma}$. For the remaining coupling constant this yields the 
constraint 
\be \label{gXschiral}
g_{\Xi\sigma} = \frac{3g_{\Lambda\sigma}-2g_{N\sigma}+g_{\Sigma\sigma}}{2} \, .
\ee
In the main part we fix all four coupling constants separately with the help of the vacuum masses, such that this relation is (slightly) violated in our phenomenological approach: with $g_{N\sigma} = 10.16$, $g_{\Lambda\sigma} = 12.07$, $g_{\Sigma\sigma} = 12.88$, the relation (\ref{gXschiral}) would yield $g_{\Xi\sigma} = 14.38$, while our fit 
gives $g_{\Xi\sigma} = 14.23$.  

Finally, for the interactions with the vector mesons, keeping only the fields $\omega$, $\rho^0$, and $\phi$, we find the structure 
\bea
&&C_1\Tr[\bar{B}\gamma^\mu V_\mu B]+C_2\Tr[\bar{B}\gamma^\mu BV_\mu]+C_3\Tr[\bar{B}\gamma^\mu B]\Tr[V_\mu]\non[2ex]
&&=
g_{N\omega}(\bar{n}\gamma^\mu\omega_\mu n+\bar{p}\gamma^\mu\omega_\mu p) 
+ g_{N\phi}(\bar{n}\gamma^\mu\phi_\mu n+\bar{p}\gamma^\mu\phi_\mu p) + g_{N\rho}(\bar{n}\gamma^\mu\rho^0_\mu n-\bar{p}\gamma^\mu\rho^0_\mu p)
\non[2ex]
&&+g_{\Sigma\omega}(\bar{\Sigma}^0\gamma^\mu\omega_\mu\Sigma^0+\bar{\Sigma}^+\gamma^\mu\omega_\mu\Sigma^++\bar{\Sigma}^-\gamma^\mu\omega_\mu\Sigma^-)
+g_{\Sigma\phi}(\bar{\Sigma}^0\gamma^\mu\phi_\mu\Sigma^0+\bar{\Sigma}^+\gamma^\mu\phi_\mu\Sigma^++\bar{\Sigma}^-\gamma^\mu\phi_\mu\Sigma^-)\non[2ex] 
&&+g_{\Sigma\rho}(\bar{\Sigma}^+\gamma^\mu\rho^0_\mu\Sigma^+-\bar{\Sigma}^-\gamma^\mu\rho^0_\mu\Sigma^-)
+g_{\Lambda\omega}\bar{\Lambda}\gamma^\mu\omega_\mu\Lambda
+g_{\Lambda\phi}\bar{\Lambda}\gamma^\mu\phi_\mu\Lambda\non[2ex]
&&+g_{\Xi\omega}(\bar{\Xi}^0\gamma^\mu\omega_\mu\Xi^0+\bar{\Xi}^-\gamma^\mu\omega_\mu\Xi^-)
+g_{\Xi\phi}(\bar{\Xi}^0\gamma^\mu\phi_\mu\Xi^0+\bar{\Xi}^-\gamma^\mu\phi_\mu\Xi^-) +g_{\Xi\rho}(\bar{\Xi}^0\gamma^\mu\rho^0_\mu\Xi^0-\bar{\Xi}^-\gamma^\mu\rho^0_\mu\Xi^-)\, .
\eea
Here, the 11 couplings are linear combinations of the 3 independent coefficients $C_1,C_2,C_3$. Equivalently, we may write $C_1,C_2,C_3$ in terms of 3 coupling constants, say the 3 nucleonic couplings $g_{N\omega}$, $g_{N\phi}$, $g_{N\rho}$, and express the remaining 8 hyperonic couplings as 
\bea
g_{\Sigma\omega} &=& \frac{g_{N\omega}+\sqrt{2}g_{N\phi}-g_{N\rho}}{2}
\, , \qquad g_{\Lambda\omega} = \frac{5g_{N\omega}+\sqrt{2}g_{N\phi}+3g_{N\rho}}{6} \, , \qquad g_{\Xi\omega} = \frac{g_{N\omega}+\sqrt{2}g_{N\phi}+g_{N\rho}}{2} \, , \non[2ex]
g_{\Sigma\rho} &=& \frac{g_{N\omega}-\sqrt{2}g_{N\phi}-g_{N\rho}}{2} \, , \qquad g_{\Xi\rho} = \frac{g_{N\omega}-\sqrt{2}g_{N\phi}+g_{N\rho}}{2} \, , \non[2ex]
g_{\Sigma\phi} &=& \frac{g_{N\omega}+g_{N\rho}}{\sqrt{2}} \, , \qquad g_{\Lambda\phi} = \frac{\sqrt{2}g_{N\omega}+4g_{N\phi}-3\sqrt{2}g_{N\rho}}{6} \, , \qquad g_{\Xi\phi} = \frac{g_{N\omega}-g_{N\rho}}{\sqrt{2}} \, .
\label{gchiral}
\eea
A particular choice for the independent coupling constants is $g_{N\phi}=0$ and $g_{N\rho}=-\frac{g_{N\omega}}{3}$. This yields the following relations,
\bea \label{gchiral2}
g_{\Sigma\omega}&=& g_{\Lambda\omega}=2g_{\Xi\omega}=\frac{2}{3}g_{N\omega} \, , \qquad  g_{\Sigma\rho}=2g_{\Xi\rho} =-2g_{N\rho} \, , \qquad 
g_{\Sigma\phi}=g_{\Lambda\phi}=\frac{g_{\Xi\phi}}{2} = \frac{\sqrt{2}}{3}g_{N\omega} \, .
\eea
These relations are often employed in the literature, see for instance Ref.\ \cite{Weissenborn:2011kb} and references therein (our sign convention for the $\rho$ and $\phi$ couplings is different compared to that reference). Also following the literature, we then fit $g_{N\omega}$ and $g_{N\rho}$ to reproduce saturation properties of nuclear matter, as explained in the main text. This violates the relation $g_{N\rho}=-\frac{g_{N\omega}}{3}$. Since this relation 
was used to derive Eqs.\ (\ref{gchiral2}) this procedure also violates the original chiral relations (\ref{gchiral}). Furthermore, we relate the $\omega$ couplings to the hyperon potential depths, ignoring the first relation of  
Eq.\ (\ref{gchiral2}). For example, for one of the parameter sets used in Sec.\ \ref{sec:4sets} we have $g_{N\omega}=10.23$, $g_{N\rho}=4.14$. With the first line of  Eqs.\ (\ref{gchiral}) this would yield $g_{\Sigma\omega}=3.05$, $g_{\Lambda\omega}=10.6$, $g_{\Xi\omega}=7.1$, while the fit to the hyperon potential ${\cal U}=-50\, {\rm MeV}$ (used for all parameter sets in Sec.\ \ref{sec:4sets}) gives the larger couplings $g_{\Sigma\omega}=14.6$, $g_{\Lambda\omega}=14.5$, $g_{\Xi\omega}=16.4$, see also Table \ref{table:para}. For the $\rho$ and $\phi$ couplings we employ the relations in Eqs.\ (\ref{gchiral2}).

\section{Asymptotic flavor symmetry}
\label{app:symmetry}

\begin{table}[t]
\hfuzz=2pt
\begin{tabular}{|c || c | c | c | } 
 \hline
    & solution 1 & solution 2 & solution 3   \\ [0.5ex] 
 \hline\hline
  \rule[-1.5ex]{0em}{6ex} 
  $g_{\Sigma\phi}$ & $g_{N\phi}+a(g_{N\omega}-g_{\Sigma\omega})$ & $g_{N\phi}+a(g_{N\omega}-g_{\Sigma\omega})$ & $\displaystyle{\frac{a^2+1}{4a}g_{\Lambda\omega}-\frac{a^2-3}{4a}g_{\Sigma\omega}}$ \\[2ex]
 \hline
 \rule[-1.5ex]{0em}{6ex} 
 $g_{\Lambda\phi}$ & $g_{N\phi}+a(g_{N\omega}-g_{\Lambda\omega}) $ &$ \displaystyle{\frac{g_{\Lambda\omega}}{a}}$ & $\displaystyle{-\frac{3a^2-1}{4a}g_{\Lambda\omega}+\frac{3(a^2+1)}{4a}g_{\Sigma\omega}}$\\[2ex] 
 \hline
 \rule[-1.5ex]{0em}{6ex} 
  $g_{\Xi\phi} $ & $\displaystyle{-\frac{3a^2-1}{a^2+1}g_{N\phi} -2a\frac{a^2-1}{a^2+1}g_{N\omega}+a\frac{3g_{\Sigma\omega}+g_{\Lambda\omega}}{2}}$ & 
  $\displaystyle{-\frac{5a^2-2}{2(a^2+1)}g_{N\phi} -a\frac{3a^2-4}{2(a^2+1)}g_{N\omega}+\frac{3a}{2}g_{\Sigma\omega}}$& $\displaystyle{-\frac{a^2-1}{a^2+1}g_{N\phi}+\frac{2a}{a^2+1}g_{N\omega}}$\\[2ex] 
 \hline
 \rule[-1.5ex]{0em}{6ex} 
  $g_{\Xi\omega}$  & $\displaystyle{\frac{4a}{a^2+1}g_{N\phi}+\frac{3a^2-1}{a^2+1}g_{N\omega}-\frac{3g_{\Sigma\omega}+g_{\Lambda\omega}}{2}}$ & 
  $\displaystyle{\frac{7a}{2(a^2+1)}g_{N\phi}+\frac{5a^2-2}{2(a^2+1)}g_{N\omega}-\frac{3}{2}g_{\Sigma\omega}}$& $\displaystyle{\frac{2a}{a^2+1}g_{N\phi}+\frac{a^2-1}{a^2+1}g_{N\omega}}$  \\[2ex]
 \hline
\end{tabular}
\caption{Three sets of conditions for the baryon-meson coupling constants, each leading to equal number densities of the three flavors at asymptotic densities for any value of the constant $a$, reproducing the behavior of asymptotically dense three-flavor QCD. Since none of the solutions seems to allow for sufficiently heavy stars they are not employed in the main part of the paper.   }
\label{table:uds}
\end{table}

In asymptotically dense three-flavor QCD, quark matter with equal numbers 
of up, down, and strange quarks is electrically neutral and beta-equilibrated. In this appendix we ask whether our model can reproduce this symmetric situation, i.e., whether there is a certain choice of parameters such that our chirally restored phase shares this property with actual quark matter. To this end, we first define the up, down, and strange number densities according to the flavor content of the baryons,
\begin{subequations} \label{nuds}
\bea
n_u &=& n_{n}+2n_{p}+n_{\Sigma^0}+2n_{\Sigma^+}+n_{\Lambda}+n_{\Xi^0}
\, , \\[2ex]
n_d &=& 2n_{n}+n_{p}+n_{\Sigma^0}+2n_{\Sigma^-}+n_{\Lambda}+n_{\Xi^-}
\, , \\[2ex]
n_s &=& n_{\Sigma^+}+n_{\Sigma^-}+n_{\Sigma^0}+n_{\Lambda}+2(n_{\Xi^-}+n_{\Xi^0})
\, . 
\eea
\end{subequations}
The condition $n_u=n_d$ together with the neutrality condition (\ref{neutral}) yields $n_e+n_\mu=0$. The solution of the stationarity equations thus has to be consistent with $\mu_e$ going to zero asymptotically. As an ansatz let us assume the following asymptotic behaviors for $\mu_n\to \infty$,
\be \label{ansatz}
\mu_e \simeq \frac{\mu_{e,\infty}}{\mu_n}\, , \qquad \sigma \simeq \frac{\sigma_\infty}{\mu_n^2} \, , \qquad 
\omega \simeq \omega_\infty\mu_n \, , \qquad 
\phi \simeq \phi_\infty\mu_n \, , \qquad 
\rho \simeq \frac{\rho_\infty}{\mu_n} \, ,
\ee
with coefficients $\mu_{e,\infty}$, $\sigma_\infty$, $\omega_\infty$, $\phi_\infty$, $\rho_\infty$ constant in the neutron chemical potential.  We shall see that this ansatz indeed leads to a valid solution of the stationarity equations, which can also be confirmed numerically. In the neutrality equation (\ref{neutral}), the only leading-order contributions proportional to $\mu_n^3$ come from $n_p$ and $n_{\Xi^-}$. Since the mass terms are of higher order due to $\sigma$ behaving like $1/\mu_n^2$, this yields the asymptotic condition $\mu_p^* = \mu_{\Xi^-}^*$. Since the $\rho$ condensate also vanishes asymptotically on account of the ansatz (\ref{ansatz}), this immediately gives the relation
\be\label{omphi}
\omega_\infty = \frac{g_{\Xi\phi} - g_{N\phi}}{g_{N\omega}-g_{\Xi\omega}}\phi_\infty \, .
\ee
Now, Eqs.\ (\ref{Eqw}) and (\ref{Eqphi}) have leading-order contributions proportional to $\mu_n^3$ which depend only on $\omega_\infty$ and $\phi_\infty$ (and none of the other coefficients of the ansatz (\ref{ansatz})). Together with Eq.\ (\ref{omphi}) these are three conditions for the two variables $\omega_\infty$ and $\phi_\infty$. Thus, in order for (\ref{ansatz}) to be a valid solution we require  (\ref{Eqw}) and (\ref{Eqphi}) to give the same condition. This can be translated into conditions for the coupling constants as follows: we insert Eq.\ (\ref{omphi}) into the leading-order contribution of Eqs.\ (\ref{Eqw}) and (\ref{Eqphi}) to eliminate $\omega_\infty$. Then, we require the four coefficients of the powers $\phi_\infty^0$, $\phi_\infty^1$, $\phi_\infty^2$, $\phi_\infty^3$ of the two equations to be identical up to a constant, say $a$, to find four conditions for the coupling constants. In fact, there are three possible solutions, i.e., three sets of four conditions, which we show in Table \ref{table:uds}. As a consistency check, one can ask whether we recover the chiral relations (\ref{gchiral}), which we would expect to reproduce flavor-symmetric matter. Indeed, solution 1 with $a=\sqrt{2}$ is satisfied by the chiral relations (\ref{gchiral}). The inverse is obviously not true: even within solution 1, since it consists of only four conditions, there are choices for the coupling constants that obey solution 1  but not the chiral relations (\ref{gchiral}) (in particular, if we allow for arbitrary values of $a$).  The solutions can be used to compute the corresponding $\phi_\infty$ and $\omega_\infty$. The results are not very instructive, but we have checked that they agree with the numerical evaluation. Similarly, one can consider the subleading contributions in $\mu_n$ to the stationarity equations to compute $\sigma_\infty$, $\mu_{e,\infty}$, $\rho_\infty$, but, again, we refrain from showing these results explicitly. The main observation is that there exist choices of the coupling constants, given by the solutions in the table, for which at asymptotically large densities $n_u=n_d=n_s$, with the flavor densities defined in Eq.\ (\ref{nuds}). However, we have not found a parameter set within the constraints of Table \ref{table:uds} which simultaneously fulfills all empirical constraints. Therefore, in the main text we content ourselves with employing parameter sets that do produce asymptotic strangeness, but not in a fraction of 1/3.

\bibliography{references}

\end{document}